\RequirePackage{fix-cm}
\PassOptionsToPackage{square,numbers,sort&compress}{natbib}
\documentclass[natbib,smallextended,hidelinks]{svjour3_arxiv} 
\bibpunct[,]{[}{]}{,}{n}{,}{,}      % onecolumn (second format)

\smartqed  % flush right qed marks, e.g. at end of proof
\usepackage{marginnote}
\usepackage{url}
\usepackage{amssymb,amsfonts,amsmath}
\usepackage{gnuplottex}
\usepackage{pifont,geometry,graphicx}
\usepackage{epsfig}
\usepackage[usenames,dvipsnames]{xcolor}
\usepackage{fancyvrb}
\usepackage{stmaryrd}
\usepackage{siunitx}
\usepackage{multirow}
\usepackage{psfrag}
\usepackage{subfig}
\usepackage{hyperref}
\hypersetup{
    colorlinks,
    linkcolor={blue!30!black},
    citecolor={blue!50!black},
    urlcolor={blue!80!black}
}
\newcommand{\bnew}[1]{{{\color{black}#1}}}
\newcommand{\jnew}[1]{{{\color{black}#1}}}
\newcommand{\bdnew}[1]{{{\color{black}#1}}}
\newcommand{\tbnew}[1]{{{\color{black}#1}}}
\newcommand{\figref}[1]{Fig.~\ref{#1}}
\newcommand{\secref}[1]{Section~\ref{#1}}

\newcommand{\tabref}[1]{Table~\ref{#1}}

\newcommand{\EE}{\begin{equation}}
\newcommand{\Ee}{\end{equation}}
\newcommand{\FF}[1]{\begin{figure}[#1]\CT}
\newcommand{\Ff}{\end{figure}}

\newcommand{\TT}[1]{\begin{table}[#1]}
\newcommand{\Tt}{\end{table}}

\newcommand{\TB}[1]{\begin{tabular}{#1}}
\newcommand{\Tb}{\end{tabular}}
\newcommand{\II}[1]{\begin{itemize}[#1]}
\newcommand{\Ii}{\end{itemize}}
\newcommand{\EN}[1]{\begin{enumerate}[#1]}
\newcommand{\En}{\end{enumerate}}

\newcommand{\CT}{\centering}

\makeatletter
\let\@afterindenttrue\@afterindentfalse
\makeatother
\AtBeginDocument{%
  \setlength{\oddsidemargin}{\dimexpr(\paperwidth-\textwidth)/2-1in}%
  \setlength{\evensidemargin}{\oddsidemargin}%
  \setlength{\topmargin}{%
    \dimexpr(\paperheight-\textheight)/2-\headheight-\headsep-1in}%
}

\newcommand{\rtwoC}[1]{\textcolor{black}{#1}}
\newcommand{\rthrC}[1]{\textcolor{black}{#1}}

\newcommand{\revC}[1]{{\color{black}{#1}}}
\newcommand{\fempty}[1]{{}}
\newcommand{\f}[1]{\mbox{$ #1 $}}

\newcommand{\sty}[1]{\mbox{\boldmath $#1$}}
\newcommand{\fa}{\sty{ a}}
\newcommand{\fb}{\sty{ b}}

\newcommand{\fd}{\sty{ d}}
\newcommand{\fe}{\sty{ e}}

\newcommand{\fl}{\sty{ l}}

\newcommand{\fn}{\sty{ n}}

\newcommand{\fr}{\sty{ r}}

\newcommand{\fG}{\sty{ G}}
\newcommand{\fH}{\sty{ H}}

\newcommand{\arcos}[1]{{\rm arcos}\left({#1}\right)}
\begin{document}
% \setcitestyle{number}
\setlength{\parindent}{0pt}
\title{\revC{Review on} Slip Transmission Criteria in Experiments and Crystal Plasticity Models%\thanks{Grants or other notes
}
% \subtitle{Do you have a subtitle?\\ If so, write it here}


\author{E.~Bayerschen \and A.T.~McBride \and B.D.~Reddy \and T.~B\"ohlke}

 %etc.


\institute{E.~Bayerschen \and T.~B\"ohlke \at
              Institute of Engineering Mechanics (ITM), Chair for Continuum Mechanics, Karlsruhe Institute of Technology (KIT), Kaiserstr.\ 10, D-76131 Karlsruhe, Germany \\
              \email{eric.bayerschen@kit.edu, thomas.boehlke@kit.edu}\\
              Tel.: +49-721-608-481-33,  +49-721-608-488-52
           \and
           A.T.~McBride \and 
	B.D.~Reddy \at
              Centre for Research in Computational and Applied Mechanics (CERECAM), University of Cape Town (UCT), 7701 Rondebosch, South Africa\\
\email{andrew.mcbride@uct.ac.za, daya.reddy@uct.ac.za}\\
\bdnew{The authors confirm that no conflicts of interest arise with regard to the research leading to this paper, nor with publication of this work.}
}

\date{The published version of this article is available in the \href{http://link.springer.com/article/10.1007/s10853-015-9553-4}{Journal of Materials Science}.}

% \url{http://link.springer.com/article/10.1007/s10853-015-9553-4}}

\maketitle

\begin{abstract}
\bdnew{A comprehensive overview is given of the literature on slip transmission criteria for grain boundaries in metals, with a focus on slip system and grain boundary orientation. Much of this extensive literature has been informed by experimental investigations.} The use of geometric criteria in continuum crystal plasticity models is discussed. The theoretical framework of Gurtin~(2008, J.~Mech.~Phys.~Solids 56, p.~640) is reviewed for the single slip case. \revC{This highlights the connections to} \rthrC{slip transmission criteria from the literature} \bnew{that are not discussed in the work itself. \rthrC{Different geometric criteria are compared for the single slip case with regard to their prediction of slip transmission.} Perspectives on additional criteria, \rthrC{investigated in experiments and} used in computational simulations, are given}.
\end{abstract}
\section{Introduction}
The plastic deformation of metals is influenced by the \rthrC{microstructure of the material, e.g., by grain boundaries (GBs)~\cite{hirth1972influence}.
%  \rthrC{Grain boundaries (GBs)~\cite{hirth1972influence} are such a characteristic, significantly affecting the plastic material response. 
Dislocations from adjacent grains interact at the grain boundary by different mechanisms~\cite{shen1988dislocation,davis1966slip,lim1985interaction,shen1986dislocation,murr1981strain,lim1984slip}. The effective transmission and activation behavior of dislocations at grain boundaries is a result of dislocation reactions~\cite{lee1989anomalous,medlin1997climb}, the type of dislocations involved~\cite{zghal2001transmission,zghal2001transmissionb}, and the microstructural characteristics such as grain boundary type~\cite{lim1985continuity,gemperlova2004slip,gemperle2005reactions,pond2006study} or phase composition~\cite{takasugi1978activated,forwood1981prismatic,de2006situ}.\\
Basic dislocation interaction and slip transfer mechanisms are illustrated in \figref{fig:transmech}.
\begin{figure}[htbp]
\begin{center}{
% \psfrag{n1}{\f{{\color{Emerald}\fn_\alpha^{\rm A}}}}
% \psfrag{d1}{\f{{\color{Emerald}\fd_\alpha^{\rm A}}}}
% \psfrag{n11}{\f{{\color{Emerald}\fn_{\alpha=1}^{\rm A}}}}
% \psfrag{d11}{\f{{\color{Emerald}\fd_{\alpha=1}^{\rm A}}}}
% \psfrag{n12}{\f{{\color{Emerald}\fn_{\alpha=2}^{\rm A}}}}
% \psfrag{d12}{\f{{\color{Emerald}\fd_{\alpha=2}^{\rm A}}}}
% \psfrag{n2}{\f{{\color{red}\fn_\beta^{\rm B}}}}
% \psfrag{d2}{\f{{\color{red}\fd_\beta^{\rm B}}}}
% \psfrag{ngb}{\f{{\color{black}\fn_\Gamma}}}
% \psfrag{d}{\f{{\color{black}\delta}}}
% \psfrag{e}{\f{{\color{black}\omega}}}
% \psfrag{k}{\f{{\color{black}\kappa}}}
% \psfrag{a}{a)}
% \psfrag{b}{b)}
% \psfrag{c}{c)}
% \psfrag{d}{d)}
% \psfrag{e}{e)}
% \psfrag{f}{f)}
% \psfrag{g}{g)}
% \psfrag{h}{h)}
% \includegraphics[width=0.95\linewidth]{Figure_01.eps}}
\includegraphics[width=0.95\linewidth]{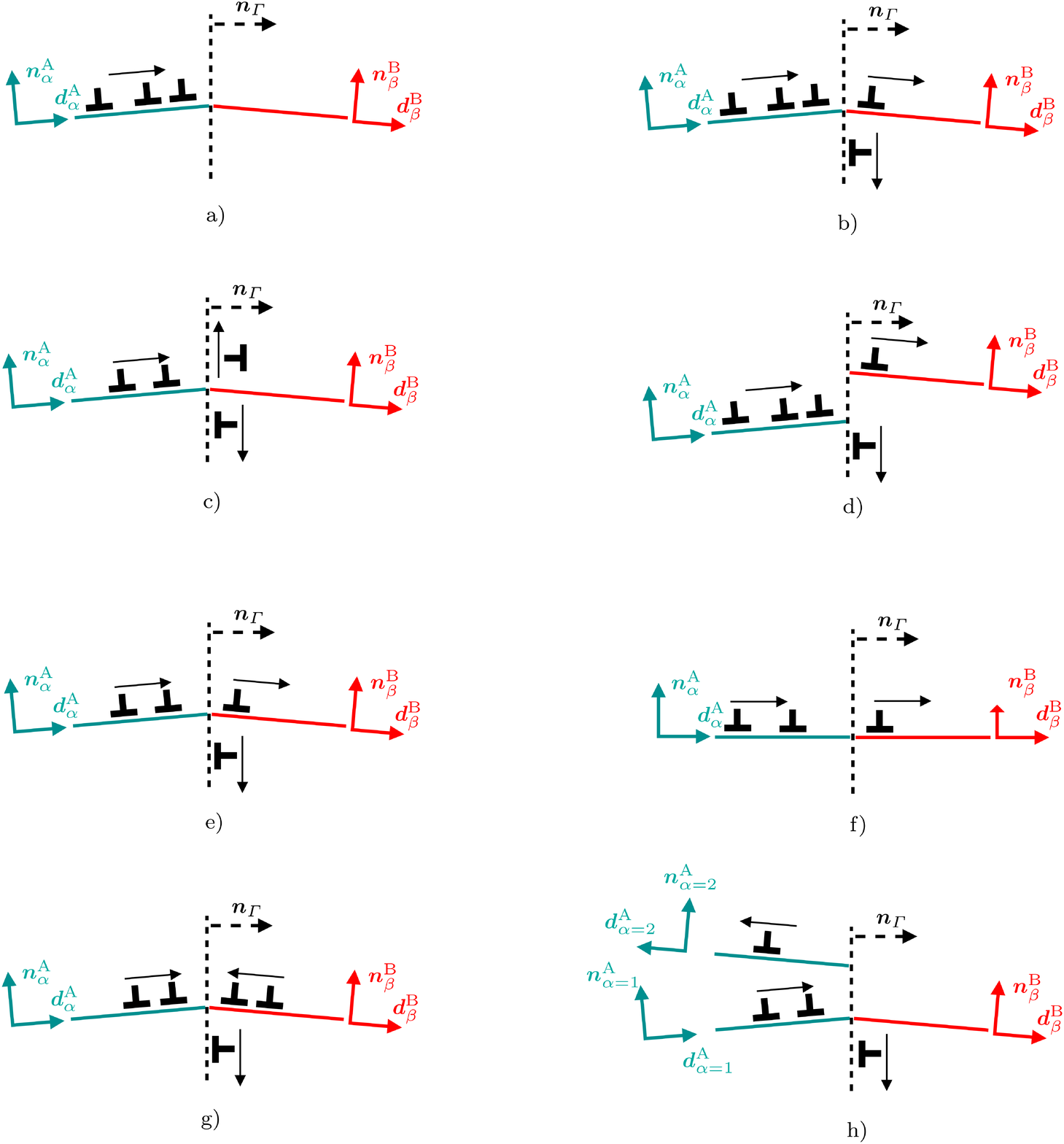}}
\end{center}
\caption{\rthrC{Slip transfer and dislocation interaction mechanisms in adjacent grains A, B, separated by grain boundary with normal~\f{\fn_\Gamma}.}}
\label{fig:transmech}
\end{figure}
The orientations of the slip directions are denoted by~\f{\fd_\alpha^{\rm A},\fd_\beta^{\rm B}}, and the orientations of the slip plane normals by~\f{\fn_\alpha^{\rm A},\fn_\beta^{\rm B}}, respectively. Here, \f{\alpha=1,\ldots,N} are slip systems of grain~A at the GB with normal~\f{\fn_\Gamma}. The number of slip systems is \f{N}, and~\f{\beta=1,\ldots,N} are the slip systems of grain~B. For example, dislocations can pile up at the grain boundary, \figref{fig:transmech}~a). Subsequently, several mechanisms are possible. Dislocation sources can be activated, leading to the emission of dislocations in grain~B and in the grain boundary, \figref{fig:transmech}~b). In addition, dislocations can be dissociated into the grain boundary, leading to no dislocation emission in grain~B, at first (\figref{fig:transmech}~c)). Afterwards, a re-emission of resolved dislocations is possible in grain~B, \figref{fig:transmech}~d). A transfer of a dislocation from the pile-up in grain~A to grain~B is depicted in \figref{fig:transmech}~e), leaving a residual dislocation in the grain boundary behind due to the required continuity of the Burgers vector~\cite{lim1985role}. For an ideal alignment of slip-directions, the dislocation is entirely transmitted and no residual dislocation is deposited in the grain boundary, \figref{fig:transmech}~f). Furthermore, mechanisms can occur involving incoming dislocations from both grains adjacent to the grain boundary like, e.g., the absorption of two dislocations generating a new dislocation in the grain boundary, \figref{fig:transmech}~g). The reflection of a dislocation from the grain boundary back into grain~A is depicted in \figref{fig:transmech}~h), leading to a residual dislocation in the grain boundary. It is remarked that the above-depicted mechanisms are simplified schematics intended for a brief overview of important mechanisms.\\
More involved mechanisms include, for example, the absorption of dislocations into the grain boundary by either dissociation of the dislocation into DSC dislocations \cite{pond1977absorption} appropriate for the specific type of grain boundary~\cite{clark1979interaction}, or by remaining as a localized dislocation. Furthermore, which of these mechanisms become active (some can also act simultaneously~\cite{shen1988dislocation}) depends not only on the lattice orientations of both grains and the grain boundary but also on the type of dislocations, i.e., whether these are edge dislocations, screw dislocations or mixed dislocations. Screw dislocations can, in principle, cross the grain boundary without leaving a residual dislocation~\cite{lim1985interaction}. There are also indications that continuous screw slip bands across a grain boundary are more likely due to a transfer of dislocations than due to an activation of dislocation sources in the adjacent grain~\cite{lim1985continuity}. Grain boundaries can also act as dislocation sources generating dislocations that remain within the vicinity of the grain boundary, and with subsequent emission into the adjacent grains~\cite{malis1979grain} (not depicted above). Dislocations can, furthermore, be generated at precipitate interfaces~\cite{brentnall1965some}.\\
Transmission of dislocations across GBs \tbnew{corresponds to the transfer of line defects and can be associated to the transfer of slip by means of the Orowan equation}. \rthrC{From an experimental point of view, the activity of slip systems in the adjacent grains to a GB can be investigated by analysis of the slip traces \cite{lall1979orientation} on both sides of the GB. However, there are various degrees of continuity or discontinuity that can be observed~\cite{seal2012analysis,west2013strain}. A discontinuous slip trace is schematically shown in \figref{fig:slip_trace}~a). In~\cite{west2013strain}, some of the developing slip traces, in the grain adjacent to a pile-up, were observed to spread over the whole grain (see schematic in \figref{fig:slip_trace}~b)), but others reached only a few microns (\figref{fig:slip_trace}~c)) into the grain. The authors of the latter work propose a classification into continuous and discontinuous slip traces, considering all grain boundaries intersected by dislocation channels. This also included grain boundaries where dislocation channels arrested on the incoming grain-side of the grain boundary, and where new channels developed a few microns away from the GB in the adjacent grain for the same slip system (illustrated in~\figref{fig:slip_trace}~d)). Furthermore, slip traces can be continuous on parts of a grain boundary and be discontinuous on other parts~\cite{bridier2005analysis,abuzaid2012slip}.
\begin{figure}[htbp]
\begin{center}{
% \psfrag{l1}{\f{{\color{Blue}\fl_\alpha^{\rm A}}}}
% \psfrag{l2}{\f{{\color{Mahogany}\fl_\beta^{\rm B}}}}
% \psfrag{n1}{\f{{\color{Emerald}\fn_\alpha^{\rm A}}}}
% \psfrag{d1}{\f{{\color{Emerald}\fd_\alpha^{\rm A}}}}
% \psfrag{n11}{\f{{\color{Emerald}\fn_{\alpha=1}^{\rm A}}}}
% \psfrag{d11}{\f{{\color{Emerald}\fd_{\alpha=1}^{\rm A}}}}
% \psfrag{n12}{\f{{\color{Emerald}\fn_{\alpha=2}^{\rm A}}}}
% \psfrag{d12}{\f{{\color{Emerald}\fd_{\alpha=2}^{\rm A}}}}
% \psfrag{n2}{\f{{\color{red}\fn_\beta^{\rm B}}}}
% \psfrag{d2}{\f{{\color{red}\fd_\beta^{\rm B}}}}
% \psfrag{ngb}{\f{{\color{black}\fn_\Gamma}}}
% \psfrag{d}{\f{{\color{black}\delta}}}
% \psfrag{e}{\f{{\color{black}\omega}}}
% \psfrag{k}{\f{{\color{black}\kappa}}}
% \psfrag{a}{a)}
% \psfrag{b}{b)}
% \psfrag{c}{c)}
% \psfrag{d}{d)}
% \psfrag{e}{e)}
% \psfrag{f}{f)}
% \psfrag{g}{g)}
% \psfrag{h}{h)}
% \includegraphics[width=0.75\linewidth]{Figure_02.eps}}
\includegraphics[width=0.75\linewidth]{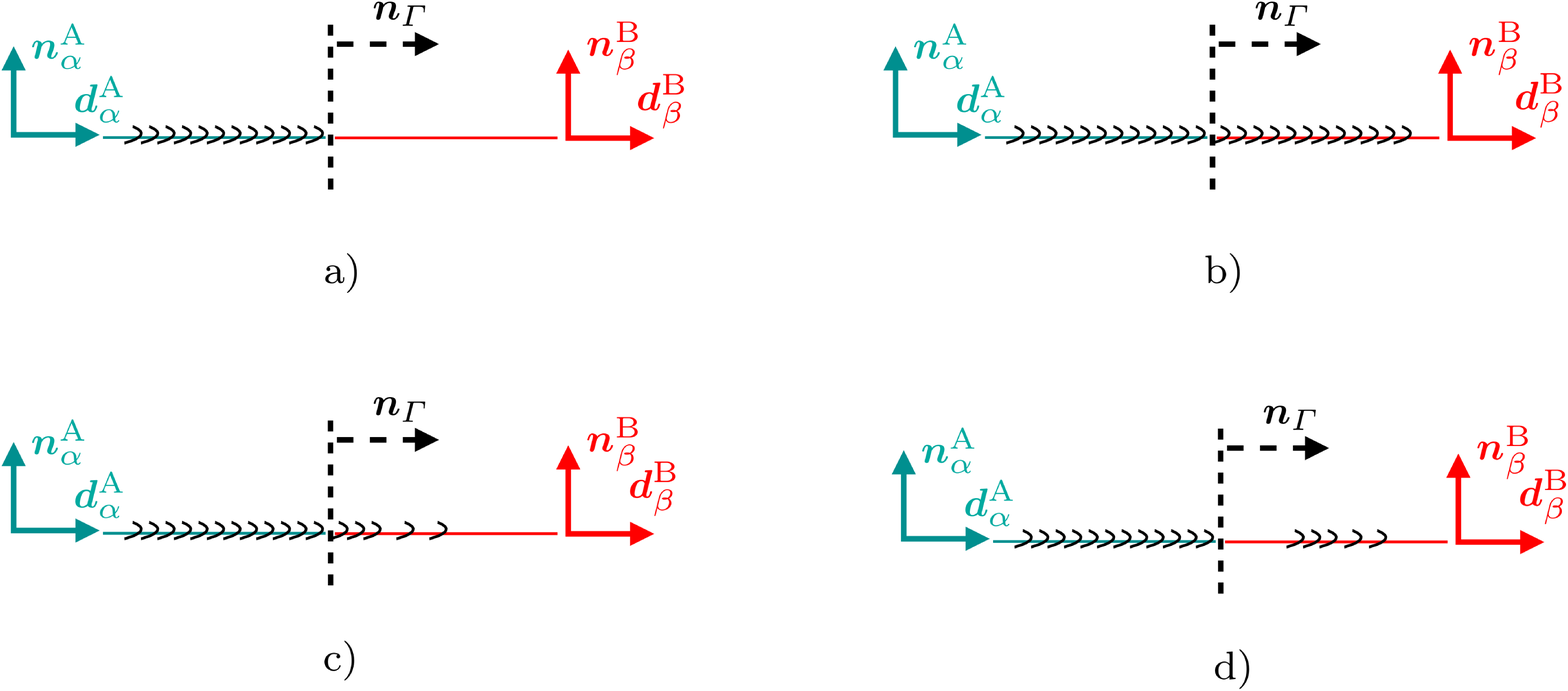}}
\end{center}
\caption{\rthrC{Discontinuity and continuity of slip traces and dislocation movement across a grain boundary with normal~\f{\fn_\Gamma} between two slip systems~\f{\alpha,\beta} in grains~A,~B.}}
\label{fig:slip_trace}
\end{figure}
}\\
\rthrC{The transfer of line defects} is influenced by both, the orientation of the GBs~\cite{shen1988dislocation}, and the orientation of the slip systems (see, e.g.,~\cite{miura1978plastic}). The better the alignment of two adjacent slip systems, the more likely a transmission event is to occur between grains~\cite{clark1992criteria}. \tbnew{Consequently, such a slip transmission criterion interconnects the slip activity on both sides of the grain boundary.} This geometric criterion seems to be the most substantial one \jnew{in predicting the correct slip system for a dislocation transmission~\cite{guo2014slip}}. Additional transmission criteria, however, have been proved to be essential, as well~\cite{lee1989prediction}. These include minimizing the residual Burgers vector (RBV) remaining in the GB upon a transmission as well as maximizing the resolved shear stress (RSS) on the outgoing slip plane. \rthrC{The RBV is the Burgers vector necessary to ensure continuity of the Burgers vector upon transmission from slip system~A to~B, i.e., \f{\fb_{\rm r}+\fb^{\rm B}=\fb^{\rm A}}~\cite{lim1985role}. Thus, the residual Burgers vector vanishes for an ideal transmission event, where both slip directions, i.e., also the Burgers vectors for both slip systems, are coinciding.} \rthrC{The three criteria of a geometric transmission factor combined with RSS and RBV have been investigated experimentally and computationally, mainly for face-centered cubic (FCC) crystal structured materials, e.g.,~\cite{kacher2012quasi,zhou2012dislocation,bachurin2010dislocation,sangid2011energy,koning2002modelling,brandl2007slip,zhu2012plastic} and often for symmetric grain boundaries~\cite{baillin1987dislocation,couzinie2005interaction,jin2006interaction,dewald2007multiscale,dewald2007multiscaleb} (see~\cite{sangid2011energy} for an overview of investigations on FCC including a list of boundary types). However, more recent works also consider general grain boundaries~\cite{bachurin2010dislocation,liu2012simulation}. In~\cite{kacher2014situ}, it was concluded that the criteria are applicable for hexagonal close-packed (HCP) materials, too. For body-centered cubic (BCC) materials, the existing work is still not exhaustive up to date, e.g.,~\cite{soer2005detection,britton2009nanoindentation,gemperlova2004slip,gemperle2005reactions}. In the latter two experimental works, the minimization of the RBV was confirmed for BCC.  There were, however, also indications that transmission criteria might be dependent on the grain boundary type. Furthermore, in the computational work~\cite{bachurin2010dislocation}, the applicability was also investigated for an asymmetric grain boundary. It was found that the inclination of the grain boundary is of minor importance for the transmission. For a detailed overview on the existing body of works regarding different crystal structures, it is referred to~\cite{kacher2014dislocation}. The criteria also appear to be applicable for interphase boundaries in alloys~\cite{misra1999slip}.} \rthrC{There are, however, also indications that the energetic structure of interphase boundaries~\cite{beyerlein2013mapping,beyerlein2014influence} needs to be considered in an extension of the criteria~\cite{beyerlein2012structure} and that additional criteria are necessary, e.g., for the case of lagging Shockley partials~\cite{dewald2007multiscale}}. \rthrC{In~\cite{misra2005length}, the strengthening mechanisms of incoherent interphase boundaries are found to be length scale dependent with a peak strength set by the interface resistance to single dislocation transmission.}}\\
% Due to continuity of the Burgers vector, a residual dislocation is deposited in the grain boundary}
\rtwoC{Continuum theories are commonly used to model the effective material response~\cite{asaro1983crystal,needleman1993comparison} of metals and to support experimental findings~\cite{yao2014plastic,ziemann2015deformation}. They offer comparably low computational time requirement and a broad spectrum of applicability to single- as well as polycrystalline materials. Applications include, for example, the prediction of hardening behavior due to plastic anisotropy~\cite{beyerlein2007plastic} and texture evolution~\cite{eyckens2015prediction}, modeling phenomena like deformation twinning~\cite{kalidindi1998incorporation}, and shear banding~\cite{forest1998modeling}. Whenever the material response is of singlecrystalline type~\cite{zaafarani2006three} or at least polycrystalline such that individual grains are not predominant~\cite{zhang2015multi}, regarding the effective material response, continuum approaches are valuable to use since the microstructure of the material does not need to be accounted for, explicitly. In cases when microstructural characteristics of the material become predominant, continuum models need to be enriched by additional considerations, like the incorporation of strain gradients~\cite{tian2014dislocation} and explicit modeling of the influence resulting from interfaces, e.g., in bimetallic materials~\cite{mayeur2015incorporating}. Modeling grain boundary mechanisms is an ongoing challenge in the development of continuum models and has been approached, e.g., by the development of gradient crystal plasticity theories, e.g.,~\cite{gurtin2008theory,wulfinghoff2013gradient,van2013grain}.} \rtwoC{It is known that five macroscopic and three microscopic degrees of freedom are necessary to specify a general grain boundary~\cite{wolf1990correlation}. Continuum models, however, commonly neglect the microscopic degrees of freedom, i.e., the translations between slip systems from adjacent grains at a grain boundary are not accounted for due to the coarsening made. Only the macroscopic degrees of freedom are then considered which define the rotations between slip systems and the grain boundary, respectively.} \jnew{\rthrC{For large-grained microstructures~\cite{kacher2014dislocation}, the GB modeling is of utmost importance} since the GB presence and influence on the dislocation movement leads to pile-ups \rthrC{of dislocations that can, in turn, dominate the material behavior.} Associated occurring phenomena are size effects, e.g.,~\cite{fleck1994strain,chen2015size}. \rtwoC{Size effects can be modeled in gradient plasticity models by different approaches. The free energy can be enhanced by terms taking into account excess dislocations~\cite{gurtin2000plasticity,gurtin2002gradient}. Thereby, gradient stresses or back-stresses are induced which enter the equations for slip rates. Such an approach neglects the influence of grain boundaries or other obstacles on the dislocation structures. If, for example, dislocation structures in dual-phase steels are to be described on the grain scale taking into account interaction of ferrit grains and the coverage of ferrit grains by martensite particles, then such an approach is insufficient~\cite{rieger2015microstructure}. The aforementioned grain interactions can be incorporated by a grain boundary yield condition~\cite{wulfinghoff2013gradient} which mimics the slip interaction on grain boundaries by additional constitutive equations which can take into account the misorientation \cite{van2013grain}, the orientation of the grain boundary relative to the grain orientations~\cite{gurtin2008theory}, and additionally the grain boundary defect energy. By this grain boundary yield mechanism, a (grain) size effect is induced.}\\
In continuum models, conservative glide of dislocations is often assumed and dislocation transfer across interfaces like GBs is modeled, \rthrC{usually, from a phenomenological perspective, e.g.,~\cite{aifantis2006interfaces}. Thereby the discrete causes of, e.g., strain fields close to grain boundaries can, however, not be distinguished anymore. For instance, continuum models incorporating grain boundary yielding can not distinguish between strain fields caused by dislocation transmission across the GB and strain fields caused by absorption of dislocations from adjacent grains into the GB~\cite{zhang2014internal}.} However, \rthrC{more sophisticated} continuum models have also been developed that incorporate \rthrC{physical} mechanisms like, for example, climbing of dislocations~\cite{geers2014coupled}. \rthrC{The model of~\cite{van2015defect} considers the redistribution of defects along the grain boundary in an averaged sense via a diffusion-type equation for the spreading of the net-defect content of the grain boundary along its planar surface. Dislocation transport is considered in the models of~\cite{reuber2014dislocation,dogge2015interface}. In the latter work, flux equations are explicitly accounted for at the interfaces, thereby modeling the transport across them. It has been experimentally determined that the changes in line energy, accompanying dislocation motion, are important in the context of grain boundary dislocations~\cite{lucadamo2002dislocation}. The framework~\cite{wulfinghoff2015gradient} takes into account the transport of dislocations in the bulk and curvature-induced line-length production by coupling a physically enriched continuum dislocation dynamics model~\cite{hochrainer2014continuum} with a simplified gradient plasticity model~\cite{wulfinghoff2013gradient}.}}\\
Existing continuum grain boundary models that account for slip transmission criteria across GBs are limited. Many models are only two-dimensional in their nature, e.g.,~\cite{evers2004scale,van2013grain}, or in their implementation~\cite{ozdemir2014modeling}. Researchers trying to incorporate slip transmission criteria in continuum models are faced with the challenge that there are several geometric slip transmission criteria, e.g.,~\cite{livingston1957multiple,luster1995compatibility,shen1986dislocation,clark1992criteria}. \bnew{Related} articles commonly include a \bnew{brief} overview on \bnew{selected works from} the experimental literature (see for example~\cite{guo2014slip,bieler2014grain,kacher2014dislocation,mayeur2015incorporating}). \bnew{A comprehensive} overview, however, that includes all geometrical slip transmission criteria in a unified and compact notation to ease the comparability of the \jnew{geometrical} concepts \bnew{is still missing in the literature to date}.\\
\revC{The outline of this work is as follows. At first, an overview is given on the geometric slip transmission criteria used in experiments and computational models including perspectives on additional criteria and considerations. Then, the sophisticated GB theory of~\cite{gurtin2008theory} (recently implemented in three dimensions in \cite{Gottschalk2016443}) is analyzed for the single slip case and the connections to the transmission criteria are discussed. Finally, the geometrical criteria are compared for the single slip case and conclusions are drawn on limitations of these geometric factors.}
% \newpage
% \begin{figure}[htbp]
% \begin{center}{
% % \psfrag{l1}{\f{{\color{Blue}\fl_\alpha^{\rm A}}}}
% % \psfrag{l2}{\f{{\color{Mahogany}\fl_\beta^{\rm B}}}}
% % \psfrag{n1}{\f{{\color{Emerald}\fn_\alpha^{\rm A}}}}
% % \psfrag{d1}{\f{{\color{Emerald}\fd_\alpha^{\rm A}}}}
% % \psfrag{n11}{\f{{\color{Emerald}\fn_{\alpha=1}^{\rm A}}}}
% % \psfrag{d11}{\f{{\color{Emerald}\fd_{\alpha=1}^{\rm A}}}}
% % \psfrag{n12}{\f{{\color{Emerald}\fn_{\alpha=2}^{\rm A}}}}
% % \psfrag{d12}{\f{{\color{Emerald}\fd_{\alpha=2}^{\rm A}}}}
% % \psfrag{n2}{\f{{\color{red}\fn_\beta^{\rm B}}}}
% % \psfrag{d2}{\f{{\color{red}\fd_\beta^{\rm B}}}}
% % \psfrag{ngb}{\f{{\color{black}\fn_\Gamma}}}
% % \psfrag{d}{\f{{\color{black}\delta}}}
% % \psfrag{e}{\f{{\color{black}\omega}}}
% % \psfrag{k}{\f{{\color{black}\kappa}}}
% % \psfrag{a}{a)}
% \psfrag{N}{b)}
% % \psfrag{c}{c)}
% % \psfrag{d}{d)}
% % \psfrag{e}{e)}
% % \psfrag{f}{f)}
% % \psfrag{g}{g)}
% % \psfrag{h}{h)}
% % \psfrag{g}{\f{\Gamma}}
% \includegraphics[width=0.3\linewidth]{Testbild_N.eps}}
% \end{center}
% \caption{\rthrC{Continuity and discontinuity of slip traces and dislocation movement across a grain boundary with normal~\f{\fn_\Gamma} between two grains~A,~B.}}
% \label{fig:N}
% \end{figure}

\section{\rthrC{Slip transmission criteria}}
\subsection{Criteria that account for slip system orientations}
\begin{figure}[htbp]
\begin{center}{
% \psfrag{l1}{\f{{\color{Blue}\fl_\alpha^{\rm A}}}}
% \psfrag{l2}{\f{{\color{Mahogany}\fl_\beta^{\rm B}}}}
% \psfrag{n1}{\f{{\color{Emerald}\fn_\alpha^{\rm A}}}}
% \psfrag{d1}{\f{{\color{Emerald}\fd_\alpha^{\rm A}}}}
% \psfrag{n2}{\f{{\color{red}\fn_\beta^{\rm B}}}}
% \psfrag{d2}{\f{{\color{red}\fd_\beta^{\rm B}}}}
% \psfrag{ngb}{\f{{\color{black}\fn_\Gamma}}}
% \psfrag{d}{\f{{\color{black}\delta}}}
% \psfrag{e}{\f{{\color{black}\omega}}}
% \psfrag{k}{\f{{\color{black}\kappa}}}
% \psfrag{GA}{Grain A}
% \psfrag{GB}{Grain B}
% \psfrag{g}{\f{\Gamma}}
% \includegraphics[width=0.7\linewidth]{Figure_03.eps}}
\includegraphics[width=0.7\linewidth]{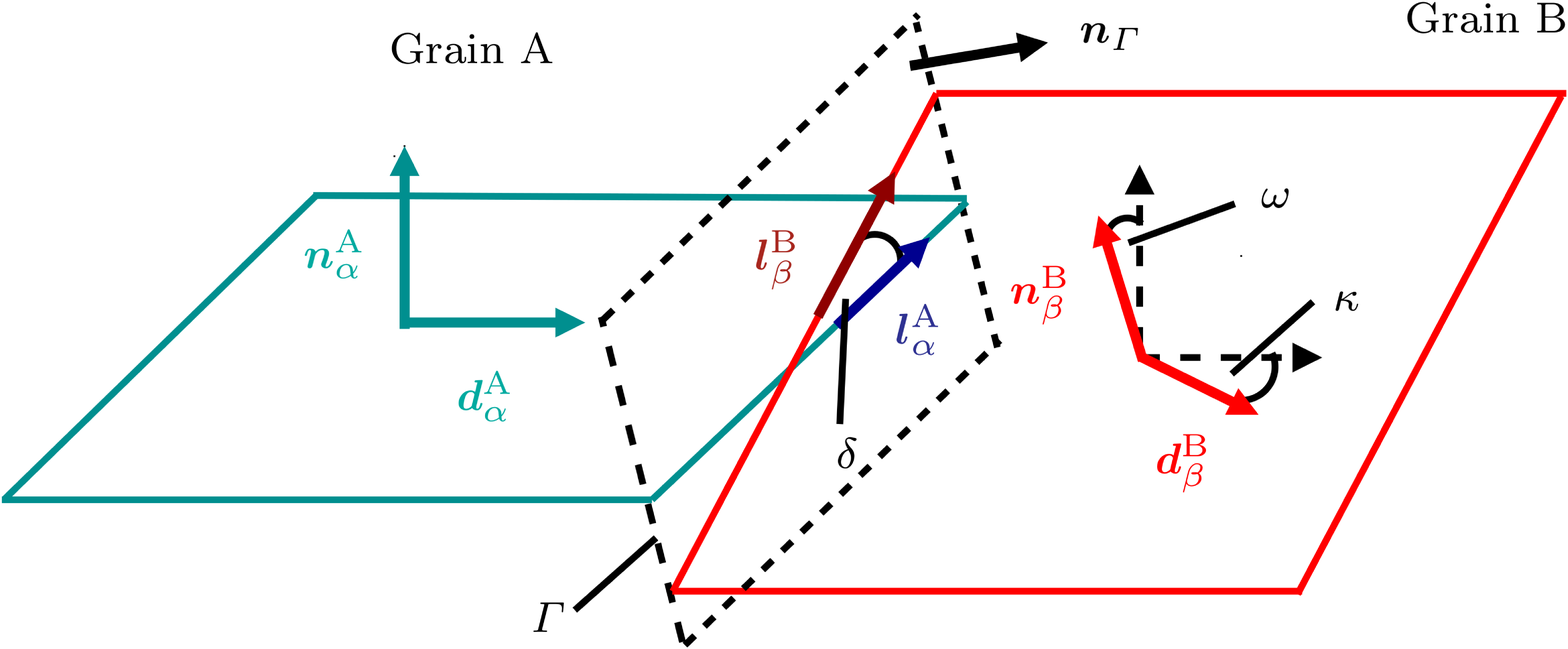}}
\end{center}
\caption{Nomenclature for slip systems~\f{\alpha,\beta} in adjacent grains A, B, separated by grain boundary~\f{\Gamma}.}
\label{fig:slipillu}
\end{figure}

Livingston and Chalmers~\cite{livingston1957multiple} were among the first to use geometric slip transmission criteria \rthrC{in experiments} to predict the activated slip system in a grain adjacent to a dislocation pile-up. Their geometric criterion accounts for the orientations of the slip directions~\f{\fd_\alpha^{\rm A},\fd_\beta^{\rm B}}, and the orientations of the slip plane normals~\f{\fn_\alpha^{\rm A},\fn_\beta^{\rm B}}, respectively (\figref{fig:slipillu}).
% transfer_mechanism.figg
The used transmission factor matrix reads
\begin{equation}
\label{eq:Nji}
 \hat N_{\alpha\beta}=(\fn_\alpha^{\rm A}\cdot\fn_\beta^{\rm B})(\fd_\alpha^{\rm A}\cdot\fd_\beta^{\rm B})+(\fn_\alpha^{\rm A}\cdot\fd_\beta^{\rm B})(\fn_\beta^{\rm B}\cdot\fd_\alpha^{\rm A}),
\end{equation}
and has \f{N\times N} components. With this criterion, the activation stress of the outgoing slip system is calculated~\cite{livingston1957multiple}. \rthrC{It is based on the approximation that the stress state in the adjacent grain, resulting from the pile-ups of the incoming slip system at the grain boundary, is of pure shear stress type.} \tbnew{Thus, the shear stresses on the incoming and on the outgoing slip systems are interconnected by the individual transmission factors.} This purely geometric criterion is also used in~\cite{davis1966slip}, \rthrC{following the interpretation of an activation of dislocation sources due to pile-ups}. A slightly modified version \rthrC{of the geometric transmission factor matrix \eqref{eq:Nji}} is employed in~\cite{luster1995compatibility}. The second term of \eqref{eq:Nji} is dropped, and the transmission factor then reads
\begin{equation} 
\label{eq:Njitilde}
{\hat N^{\rm mod}}_{\alpha\beta}=(\fn_\alpha^{\rm A}\cdot\fn_\beta^{\rm B})(\fd_\alpha^{\rm A}\cdot\fd_\beta^{\rm B}).
\end{equation}
This factor is combined with the Schmid factors~\cite{schmid1935kristallplastizitat} and a stress intensity factor resulting from pile-ups (based on~\cite{eshelby1951xli}) in the transmission evaluation of~\cite{guo2014slip}. It was found that a lower stress intensity factor (leading to a lower RSS) on the emission slip system correlated to larger RBVs. For controlling the slip system activation, good alignment of slip systems has proved to be more important than a high Schmid factor. \rthrC{The importance of misalignment of slip systems for slip transmission processes has also been demonstrated with the preceding geometric factor in micro-hardness measurements of GBs \cite{wo2004investigation}.}
\subsection{Criteria that account for slip system orientations and grain boundary orientation}
\begin{figure}[htbp]
\begin{center}{
\includegraphics[width=0.95\linewidth]{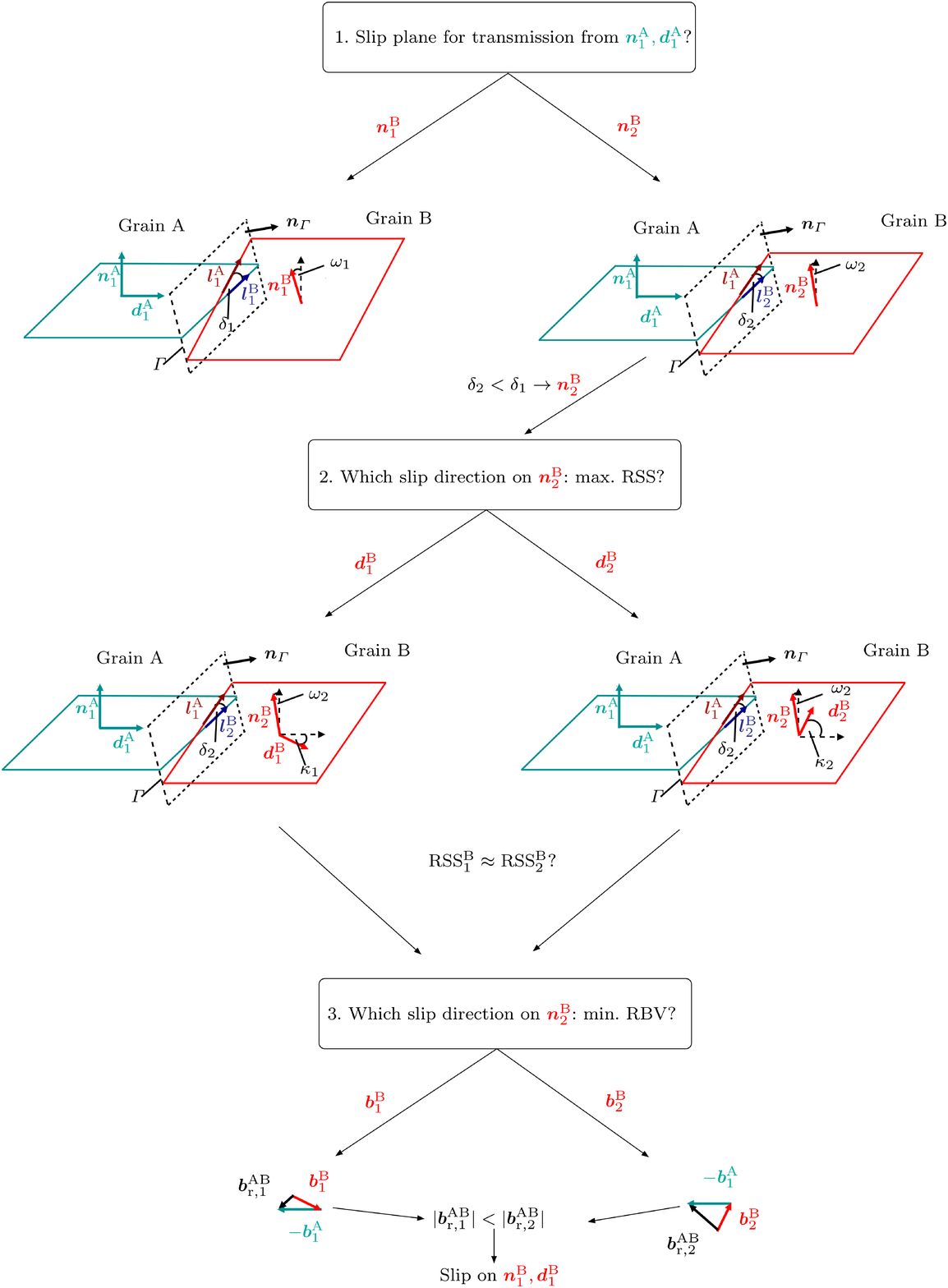}}
\end{center}
\caption{\rthrC{Combined criteria~\cite{lee1989prediction} for slip transmission from grain A to B, separated by grain boundary~\f{\Gamma}.}}
\label{fig:slipcriteria}
\end{figure}

In~\cite{shen1986dislocation}, using \eqref{eq:Nji} is compared to a different transmission factor incorporating the grain boundary orientation via
\begin{equation}
\label{eq:Mji}
   \hat M_{\alpha\beta}=(\fl_\alpha^{\rm A}\cdot\fl_\beta^{\rm B})(\fd_\alpha^{\rm A}\cdot\fd_\beta^{\rm B}),
\end{equation}
\revC{where \f{\fl_\alpha^{\rm A},\fl_\beta^{\rm B}} \revC{are normalized vectors of the} lines of intersection, see \figref{fig:slipillu}. They can be obtained from, e.g., \f{\fl_\alpha^{\rm A}=({\fn_\alpha^{\rm A}\times\fn_\Gamma})/|({\fn_\alpha^{\rm A}\times\fn_\Gamma})|}.} Here, \f{\fn_\Gamma} denotes the GB normal. In combination with a stress criterion based on \rthrC{maximizing} the Peach-Koehler force \rthrC{on the emitted dislocation}, \eqref{eq:Mji} was shown to successfully predict all slip system activations, whereas the purely geometric criterion \eqref{eq:Nji} did not. \rthrC{The geometric criterion \eqref{eq:Mji} determines the slip plane, and the stress criterion determines the slip direction of the emitted dislocation in this procedure.} These criteria were also used to predict the activation of slip systems in~\cite{shen1988dislocation}.\\
The criteria for slip transmission were further extended in~\cite{lee1989prediction} to account for the RBV, where
\begin{equation}
\label{eq:Mjitilde}
 {\hat M^{\rm mod}}_{\alpha\beta} =\fl_\alpha^{\rm A}\cdot\fl_\beta^{\rm B}
\end{equation}
was used instead of \eqref{eq:Mji}. \rthrC{In a first step, the slip plane for a possible transmission is determined by finding the slip plane normal that maximizes the scalar product between the lines of intersection. This corresponds to a minimization of the angle~\f{\delta}, see ``1.'' in \figref{fig:slipcriteria}.}
 \rthrC{Then, the slip direction on the given slip plane is determined by finding the maximum RSS on the outgoing slip directions (2.\ in \figref{fig:slipcriteria}). In case of multiple slip directions with similar RSS, the direction is chosen which minimizes the RBV~\f{\fb_{\rm r}} left behind in the GB by a transmission event. Thus, the angle~\f{\kappa} between the slip directions is minimized (3.\ in \figref{fig:slipcriteria}).} This approach removed remaining inconsistencies highlighted in the approach of~\cite{shen1988dislocation}, and was used by~\cite{abuzaid2012slip}.\\
In~\cite{lee1990tem}, it is proposed that the criteria of maximum RSS and minimum RBV need to be combined as they are competitive in nature (see also~\cite{lim1985interaction}). It was found, however, that minimizing the RBV is of dominant influence for the slip transmission. In~\cite{clark1992criteria}, it is outlined, with reference to~\cite{bamford1988thermodynamic}, that this combined criterion is not applicable to multiple active slip systems. \rthrC{For the case of intermetallic phase boundaries, the combined criteria by \cite{lee1990tem} were confirmed to be applicable with a slight refinement~\cite{misra1999slip}. This refinement considers metals with multiple types of slip systems.} The purely geometric criterion \eqref{eq:Mji} is applied in~\cite{soer2005detection}, while in~\cite{tiba2015incompatibility}, it is both applied and combined with investigations regarding the RBV criterion and incompatibility stresses. The importance of considering the RBV in the slip transmission prediction is emphasized in~\cite{patriarca2013slip}, as well.
\subsection{Criteria that consider threshold values for the slip system and grain boundary angles}
\rthrC{The geometrical criteria outlined in the previous sections were evaluated for each slip system, individually, to determine the most likely system for slip transfer across grain boundaries. However, it has also been proposed in the literature to calculate an overall measure of slip transfer by, e.g., summing over the individual components of all possible slip system combinations. Such an approach is used in~\cite{werner1990slip}.} The mismatch between slip systems in adjacent grains is \rthrC{taken into account} via a sum of the form
\begin{equation}
\label{eq:lambda}
  \lambda=\sum\limits_{\alpha=1}^{N}\sum\limits_{\beta=1}^{N}\cos\left(\frac{\ang{90}}{\rthrC{\omega}_{\rm c}}\arcos{\fn_\alpha^{\rm A}\cdot\fn_\beta^{\rm B}}\right)\cos\left(\frac{\ang{90}}{\kappa_{\rm c}}\arcos{\fd_\alpha^{\rm A}\cdot\fd_\beta^{\rm B}}\right).
\end{equation}
\rthrC{Here, \f{\omega_{\rm c}} and~\f{\kappa_{\rm c}} are critical angles above which no slip transfer is expected to occur on the associated slip systems. Thus, these slip system combinations are not considered and removed from the sum if their shared angle is too large.} \rthrC{As it is discussed in \cite{werner1990slip},} the mismatch between slip plane normals was taken into account, rather than the mismatch between lines of intersection on~\f{\Gamma}. The GB orientation was difficult to measure. The critical angles, above which slip transmission is not expected to occur, are taken to be~\f{\kappa_{\rm c}=\ang{45}} and~\f{\rthrC{\omega}_{\rm c}=\ang{15}} \revC{for $\alpha/\alpha$- and $\alpha/\beta$-phase boundaries in brass. For $\beta/\beta$-phase boundaries \f{\rthrC{\omega}_{\rm c}=\ang{30}} is used as critical angle for the slip plane normals}. The limit angle \revC{\f{\rthrC{\omega}_{\rm c}=\ang{15}}} was motivated by the work of~\cite{davis1966slip} using \eqref{eq:Nji}, where the critical angle~\f{\delta_{\rm c}} between the lines of intersection is estimated to be in the range of~\f{\ang{10}-\ang{20}}. Furthermore, it is argued that the angle of lines of intersection for a pair of slip systems on adjacent sides of a GB cannot exceed the angle between adjacent slip plane normals, i.e., \f{\delta\leq\rthrC{\omega}\rightarrow\fl_\alpha^{\rm A}\cdot\fl_\beta^{\rm B}\leq\fn_\alpha^{\rm A}\cdot\fn_\beta^{\rm B}}. Thus, \f{\rthrC{\omega}} is used in place of \f{\delta}, see \figref{fig:slipillu}. The approach of~\cite{werner1990slip} confirmed the experimental behavior of phase- / grain boundaries with regard to their slip permeability.\\
In~\cite{kumar2010influence}, however, \eqref{eq:lambda} was used in combination with the Schmid factors to investigate both criteria regarding the tensile strength of the considered material. It was found that the trend of the tensile strength was opposite to that of the calculated transmission number~\f{\lambda}, i.e., a high value of~\f{\lambda} did not lead to an increased yield strength.\\
In~\cite{beyerlein2012structure}, the angle~\f{\delta} between the lines of intersection is taken into account, rather than the angle~\f{\rthrC{\omega}} between the slip plane normals. Furthermore, instead of the summation in \eqref{eq:lambda}, individual components
\begin{equation}
\label{eq:chi}
 \hat\chi_{\alpha\beta}=\cos\left(\frac{\ang{90}}{\delta_{\rm c}}\arcos{\fl_\alpha^{\rm A}\cdot\fl_\beta^{\rm B}}\right)\cos\left(\frac{\ang{90}}{\kappa_{\rm c}}\arcos{\fd_\alpha^{\rm A}\cdot\fd_\beta^{\rm B}}\right)
\end{equation}
are considered. The same critical angles as in the previous works, however, are utilized. The geometrical criterion was combined with the Schmid factors and further considerations regarding the interface shear strength \rthrC{\cite{demkowicz2011structure,wang2011influence,wang2012structure}}.
\subsection{Criteria that consider weighted sums of geometric transmission factors}
\label{subsec:weightedsums}
\rthrC{Besides \eqref{eq:lambda}, other summation approaches exist in the literature, additionally, considering weights for the slip system contributions. Such an approach is taken in~\cite{bieler2014grain} since no clear correspondence to the transmission events could be established using only the geometric factor \eqref{eq:Njitilde}.} Several weighted sum approaches for a slip transmission factor are proposed in~\cite{bieler2014grain}. These scalar measures are based on the above described geometric factors. They are obtained by the summation over all slip system transmission factors and weighting each one with plastic slips~\f{\gamma_\alpha^{\rm A}} or the Schmid factors~\f{m^{\rm A}_\alpha,m^{\rm B}_\beta}. Two such criteria are proposed in \cite{bieler2014grain} by
\begin{subequations}
  \begin{align}\label{eq:mgamma}
m_m'&=\sum\limits_{\alpha,\beta}{\hat N^{\rm mod}}_{\alpha\beta}m_\alpha^{\rm A} m_\beta^{\rm B}\Big/\sum\limits_{\alpha,\beta}m_\alpha^{\rm A} m_\beta^{\rm B},\\
  m_\gamma'&=\sum\limits_{\alpha,\beta}{\hat N^{\rm mod}}_{\alpha\beta}\gamma_\alpha^{\rm A}\gamma_\beta^{\rm B}\Big/\sum\limits_{\alpha,\beta}\gamma_\alpha^{\rm A}\gamma_\beta^{\rm B}.
  \end{align}
\end{subequations}
% \begin{equation}
% \label{eq:mgamma}
%   m_\gamma'=\sum\limits_{\alpha,\beta}{\hat N^{\rm mod}}_{\alpha\beta}\gamma_\alpha^{\rm A}\gamma_\beta^{\rm B}\Big/\sum\limits_{\alpha,\beta}\gamma_\alpha^{\rm A}\gamma_\beta^{\rm B},\quad m_m'=\sum\limits_{\alpha,\beta}{\hat N^{\rm mod}}_{\alpha\beta}m_\alpha^{\rm A} m_\beta^{\rm B}\Big/\sum\limits_{\alpha,\beta}m_\alpha^{\rm A} m_\beta^{\rm B}.
% \end{equation}
\revC{Measure}~\rthrC{\eqref{eq:mgamma}} connects the geometric mismatch with the RSS due to the employed weighting using Schmid factors. \revC{Two other measures are also given in~\cite{bieler2014grain}} by
% \begin{equation}
\begin{subequations}
  \begin{align}
\label{eq:LRBgamma}
 s_\gamma&=\sum\limits_{\alpha,\beta}{\hat M^{\rm mod}}_{\alpha\beta}{\hat N^{\rm mod}}_{\alpha\beta}\gamma_\alpha^{\rm A}\gamma_\beta^{\rm B}\Big/\sum\limits_{\alpha,\beta}\gamma_\alpha^{\rm A}\gamma_\beta^{\rm B},\\
 LRB_\gamma&=\sum\limits_{\alpha,\beta}\hat{M}_{\alpha\beta}\gamma_\alpha^{\rm A}\gamma_\beta^{\rm B}\Big/\sum\limits_{\alpha,\beta}\gamma_\alpha^{\rm A}\gamma_\beta^{\rm B}.
  \end{align}
\end{subequations}
For the sample investigated in~\cite{bieler2014grain}, all four measures give similar distributions along the grain boundaries. \rthrC{This raises the question if the weighting by resolved shear stresses or by plastic slips is applicable. Furthermore, in \eqref{eq:LRBgamma} the slip system normals seem to be double-accounted for due to the combination of the geometric factors \eqref{eq:Njitilde} and \eqref{eq:Mjitilde}.}\\
The slip transmission criteria used in experiments are summarized in \tabref{tab:appa}.
\begin{table}[htbp]
\caption{Slip transmission criteria in the experimental literature. Abbreviations used are RSS: Resolved shear stress / Schmid factors, RBV: Residual Burgers vector, PKF: Peach-Koehler force, IBS: Interface barrier strength, SIF: Stress intensity factor, SW: Slip weights, SFW: Schmid factor weights.}
\centering
{
\def\arraystretch{1.25}
{\begin{tabular}{||c|c||c|c|c|c||}
\hline
  {Transmission} & {Additional criteria}&TF&\multicolumn{3}{c||}{Additional criteria}\\
    factor (TF)&Reference&  &\multicolumn{3}{c||}{Reference}\\\hline
 \multirow{2}{*}{\f{\hat{N}_{\alpha\beta}}}&-&\multirow{2}{*}{\f{{\hat N^{\rm mod}}_{\alpha\beta}}}&-&RSS / RBV / SIF&RSS\\
 &\cite{livingston1957multiple,davis1966slip}&&\cite{luster1995compatibility,wo2004investigation}&\cite{guo2014slip}&\cite{bieler2014grain}\\\hline
  \multirow{2}{*}{\f{{\hat M^{\rm mod}}_{\alpha\beta}}}&RSS / RBV&\multirow{2}{*}{\f{\hat{M}_{\alpha\beta}}}&-&PKF&RBV\\
  &\cite{clark1992criteria,lee1989prediction,lee1990tem,misra1999slip,abuzaid2012slip}&&\cite{soer2005detection}&{\cite{shen1986dislocation,shen1988dislocation}}&\cite{tiba2015incompatibility}\\\hline
   \multirow{2}{*}{\f{\hat\chi_{\alpha\beta}}}&RSS / IBS&\multirow{2}{*}{\f{\lambda}}&-&\multicolumn{2}{c||}{RSS}\\
   &\cite{beyerlein2012structure}&&\cite{werner1990slip}&\multicolumn{2}{c||}{\cite{kumar2010influence}}\\\hline
   \multirow{2}{*}{\f{m_\gamma'}, \f{m_m'}}&SW, SFW&\multirow{2}{*}{\f{LRB_\gamma}, \f{s_\gamma}}&\multicolumn{3}{c||}{SW, SW}\\
   &\cite{bieler2014grain}&&\multicolumn{3}{c||}{\cite{bieler2014grain}}\\\hline
\end{tabular}
}}
\label{tab:appa}
\end{table}
% \newpage

\section{Computational modeling}
\subsection{\rthrC{Computational investigation of slip transmission criteria}}
\rthrC{The criteria} of combining a geometric transmission factor (GTF), RSS, and RBV were \revC{investigated} in atomistic simulations (see~\cite{spearot2014insights} and~\cite{bieler2009role} for an overview) and molecular dynamics simulations~\cite{koning2002modelling}. \rthrC{However, it has been found that the local energetic structure and the local stress state of the GB can not be neglected in the context of dislocation interactions near GBs, in general~\cite{spearot2014insights}. Additional considerations include, e.g., the interface shear strength~\cite{demkowicz2011structure,wang2011influence,wang2012structure}.} In the atomistic simulations~\cite{sangid2012energetics} and in the combined computational / experimental approaches~\cite{abuzaid2012slip}, the importance of the RBV for the slip transmission has been demonstrated. \rthrC{The barrier provided for dislocation motion by two twist and tilt grain boundaries, respectively, was found to be proportional to the RBV magnitude. This was also the case for an investigated twin boundary.} The coupled atomistic / discrete dislocation framework~\cite{dewald2007multiscale} also confirms the \rthrC{three basic slip transmission criteria (GTF, RSS, RBV)} \rthrC{for a tilt grain boundary impinged by edge dislocations.} It is proposed there, however, that additional criteria are necessary for the case of GB dislocation nucleation. \rthrC{For the case of screw dislocations impinging on the same grain boundary type (and other symmetric tilt boundaries), no transmission but only nucleation was observed in~\cite{dewald2007multiscaleb}. In~\cite{dewald2011multiscale}, dislocations of mixed character were investigated for the same types of boundary as in~\cite{dewald2007multiscaleb}. The set of criteria outlined in~\cite{dewald2007multiscaleb} were extended since the effects of local GB structure are not accounted for by the classic criteria. They were incorporated additionally by quantitative criteria~\cite{dewald2011multiscale}.\\
By physically detailed simulation approaches like atomistics, the interaction of dislocations, e.g., with twin boundaries~\cite{ezaz2011energy}, can be investigated very thoroughly. However, contrary to many continuum models, immense computational costs arise due to the detailed modeling of interactions and the inherent discreteness of the models. This limits using discrete models for larger structures and necessitates the development of, e.g., mesoscale approaches like crystal plasticity models.}
\subsection{Crystal plasticity models taking into account geometrical slip transmission criteria}
\label{sec:transmission_cm}
The previously described criteria \revC{allow for the evaluation of slip transmission prediction by} dislocation based crystal plasticity models~\cite{zikry1996inelastic}. \revC{They can also be} explicitly incorporated in continuum models to account for the transmission mechanisms. \rtwoC{Continuum models, however, lack the discreteness inherent to simulation approaches like discrete dislocation dynamics or molecular dynamics. Therefore, the incorporation of dislocation transmission and activation processes near GBs can only be performed in an averaged sense.} In the model of~\cite{ekh2011influence}, for example, a functional relationship is proposed for the GB (slip transmission) strength. The strength depends on the minimum angle between the slip directions of slip systems in adjacent grains via
\begin{equation}\label{eq:tan}
 \tan{(\varphi_\alpha^{\rm AB})}=\tan{(\min_{\beta}(\arcos{|\fd_\alpha^{\rm A}\cdot\fd_\beta^{\rm B}|}))}.
\end{equation}
The higher the minimum angle~\f{\varphi_\alpha^{\rm AB}}, the higher is the GB strength. This criterion, however, does consider the orientations of the GB normal and slip plane normals. \revC{The resolved shear stresses are accounted for in the flow rule for the slip systems. Thereby, it is ensured that slip systems with large resolved shear stresses do yield while others with lower resolved shear stresses do not.}\\
In~\cite{shi2009grain,shi2011modeling},~\eqref{eq:Mji} is utilized in combination with \revC{a} RSS criterion. \rthrC{In case the geometric transmission factor is greater than a critical threshold, and if the ratio of resolved shear stress of an outgoing system with respect to a (with dislocation density evolving) reference shear stress is larger than one, dislocation density can pass the GB in this model and increase the density in the adjacent grain.} The thermally activated transmission approach of~\cite{ma2006consideration} assumes that the slip lines of dislocations align with the GB during transmission. They propose a criterion that is based on the minimization of the energy for a transmission event. This incorporates the RBV in the GB as well as the slip system and GB orientation. \rtwoC{In the employed flow rule, the RSSes are considered and a cutting stress is calculated which models forest dislocations as well as the GB activation energy barrier. Thus, the flow rule connects the minimization of the RBV upon transmission with the maximization of RSSes.}\\
% The orientation dependent activation energy enters the flow rule and thus connects the orientation dependence and RBV with the RSSes.\\
The GB model~\cite{gurtin2008theory} has \rtwoC{been implemented within a two-dimensional setting in~\cite{ozdemir2014modeling} and, recently, also within three dimensions~\cite{Gottschalk2016443}}. In this model, so-called {\it inter}-action coefficients describe the interaction of slip systems in adjacent grains,
\begin{equation}
\label{eq:CABG}
 \hat C^{AB}_{\alpha\beta}=(\fd_\alpha^{\rm A}\cdot\fd_\beta^{\rm B})(\fl_\alpha^{\rm A}\cdot\fl_\beta^{\rm B}).
\end{equation}
In fact, the {\it inter}-action coefficients in~\eqref{eq:CABG} are \revC{formally identical} to the geometric slip transmission factor~\eqref{eq:Mji}. The model in~\cite{gurtin2008theory}, furthermore, accounts for the RBV criterion and the RSS criterion, as well (see \secref{sec:residualB}). The superscripts \{A,B\} distinguish the {\it inter}-action coefficients from the so-called {\it intra}-action coefficients. The {\it intra}-action coefficients determine the interaction of slip systems within each grain based on~\eqref{eq:CABG}, applied to each grain \{A,B\}, individually. They read \f{\hat C^{AA}_{\alpha\beta}}, and \f{\hat C^{BB}_{\alpha\beta}}, respectively.
\subsection{Criteria that consider threshold values for the slip system and grain boundary angles}
In~\cite{ashmawi2002prediction},~\eqref{eq:Mjitilde} is extended to account for the slip plane normals intersection angle via an additional term
\begin{equation}\label{eq:zeta}
\hat\zeta_{\alpha\beta}=(\fl_\alpha^{\rm A}\cdot\fl_\beta^{\rm B})(\fn_\alpha^{\rm A}\cdot\fn_\beta^{\rm B}).
\end{equation}
Critical angles (motivated by~\cite{davis1966slip} and~\cite{werner1990slip}) are used with~\f{\rthrC{\omega}_{\rm c}=\ang{35}} and \f{\delta_{\rm c}=\ang{15}}. The employed slip transmission \revC{factor} is purely geometric, but it is combined with the dislocation densities and their evolution on the adjacent sides of GBs. \rtwoC{The RSSes are considered in the flow rule. For determining a possible transmission of dislocation density across the GB, the signs of the slip rates are checked, i.e., it is determined whether dislocations in a pile-up move towards the GB or away from it. Thereby, the geometric factor is connected to the RSSes.}\\
In the work of~\cite{mayeur2015incorporating},~\eqref{eq:chi} is used to penalize slip transfer on geometrically unfavorable slip system combinations across bimetallic interfaces by increasing the corresponding slip resistances depending on the mismatch. The modified slip resistance enters the flow rule and, thus, connects the geometrical factors to the RSSes \rtwoC{in the flow rule}.
Slip transmission criteria in continuum models are summarized in \tabref{tab:appb}.
{\def\arraystretch{1.5}
\begin{table}[htbp]
\caption{Slip transmission / interaction criteria in crystal plasticity models. Abbreviations used are RSS: Resolved shear stress / Schmid factors, RBV: Residual Burgers vector. The approaches by~\cite{gurtin2008theory,ozdemir2014modeling} utilize \f{\hat M_{\alpha\beta}} rather as an {\it inter}-action coefficient than as a classic transmission factor.}
\centering
{
\begin{tabular}{||c|c|c|c|c|c|c||}
\hline
  {Transmission factor} &\f{ \tan{(\varphi_\alpha^{\rm AB})}} & \f{\hat M_{\alpha\beta}} &\f{\hat M_{\alpha\beta}}& - &\f{\hat\zeta_{\alpha\beta}}&\f{\hat\chi_{\alpha\beta}}\\
  Additional criteria &\revC{RSS}&RSS&RBV / RSS&RBV / RSS&\revC{RSS}&RSS\\
  Reference&\cite{ekh2011influence}&\cite{shi2009grain,shi2011modeling}&\cite{gurtin2008theory,ozdemir2014modeling}&\cite{ma2006consideration} &\cite{ashmawi2002prediction}&\cite{mayeur2015incorporating}\\\hline
\end{tabular}
}
\label{tab:appb}
\end{table}}

\section{A connection between Gurtin's grain boundary theory and \rthrC{slip transmission criteria used in experiments}}
\label{sec:residualB}
\rthrC{The single-crystal plasticity framework of \cite{gurtin2002gradient} uses the measure of a Burgers tensor field to characterize the Burgers vectors of geometrically necessary dislocations. This measure is defined by~\f{\fG={\rm curl}(\fH^{\rm p})} for the geometrically linear case, where~\f{\fH^{\rm p}=\sum_\alpha\gamma_\alpha\fd_\alpha\otimes\fn_\alpha} is the plastic distortion. Precisely, the defect contribution to the free energy is formulated in dependence of this quantity. The need for a defect contribution to the free energy in continuum models results from the coarsening error made by the continuum modeling of the elastic energy \cite{mesarovic2010plasticity}. By considering the Burgers tensor in the free energy, dependences of the related higher-order stresses on this measure can be introduced on the individual slip systems, subsequently. These stresses are considered in the flow rule for the slip systems and, thus, influence the model response.\\
In \cite{gurtin2005boundary}, the theory has been extended by consideration of interfaces such as grain boundaries. The associated boundary conditions are prescribed by the limits of microfree and microhard conditions. Subsequently, the framework \cite{gurtin2008theory} has been developed incorporating the misorientation of adjacent grains and its influence on the slip transfer behavior at the grain boundaries. This is accomplished by considering the Burgers tensor field on the grain boundary.}
\rthrC{The magnitude} \f{|\fG|} is used as a measure of defect in the GB free energy. From the GB energy, internal (energetic) microforces can be derived. These, in turn, are balanced on the GB with the projections of the vectors of gradient stresses from each grain. Furthermore, these gradient stresses enter a microforce balance for each slip system~\f{\alpha} in which the RSSes enter \revC{as well}.\\
% The continuum crystal plasticity framework by \cite{gurtin2008theory} is based on the single-crystal plasticity framework by \cite{gurtin2005boundary}. The latter accounts for boundary conditions at interfaces such as GBs and 
\bnew{Although one might expect Gurtin's theory of grain boundaries~\cite{gurtin2008theory} to be connected to criteria of slip system interaction that have been used \rthrC{in experiments}, the framework used in the mentioned work is not discussed from this point of view.} \bnew{Therefore,} a single slip case is considered \bnew{in the work at hand} showing {the connections} between~\cite{gurtin2008theory} and \rthrC{the criteria} of GTF / RSS / RBV. For convenience, in the following, the single slip systems in grain~A and B are labeled A and B, respectively. For brevity, the slip plane normals of the two slip systems on adjacent sides of the GB are considered to be coinciding, i.e., \f{\fn^{\rm A}=\fn^{\rm B}=\fn}, {and to be perpendicular to the GB normal~\f{\fn_\Gamma}.} Thus, the angles~\f{\delta=\rthrC{\omega}=0}, while~\f{\kappa\neq0}, see also Fig.~\ref{fig:slipillu}. The RBV can be defined as the difference between the Burgers vectors of {interacting, i.e., transmitting} slip systems, \f{\fb_{\rm r}+\fb^{\rm B}=\fb^{\rm A}}~\cite{lim1985role}.
{Its magnitude can be approximated by the magnitude of the difference between the two slip directions~\f{\fd^{\rm A},\fd^{\rm B}}, i.e., \f{|\fb_{\rm r}|=|\fd^{\rm A}-\fd^{\rm B}|}~\cite{abuzaid2012slip}.}
Furthermore, the definition~\cite{gurtin2008theory} of the jump of the plastic distortion~\f{\fH^{\rm p}} across the GB is considered. For the single slip transmission case at hand, this jump reads
\begin{equation}
 \llbracket\fH^{\rm p}\rrbracket=\gamma^{\rm B}\fd^{\rm B}\otimes\fn-\gamma^{\rm A}\fd^{\rm A}\otimes\fn=(\gamma^{\rm B}\fd^{\rm B}-\gamma^{\rm A}\fd^{\rm A})\otimes\fn.%=-\fb_{\rm r}\otimes\fn.
\end{equation}
This gives a GB Burgers tensor~\f{\fG}~\cite{gurtin2008theory} of
\begin{equation}
 \fG=(\gamma^{\rm B}\fd^{\rm B}-\gamma^{\rm A}\fd^{\rm A})\otimes(\fn\times\fn_\Gamma)=(\gamma^{\rm B}\fd^{\rm B}-\gamma^{\rm A}\fd^{\rm A})\otimes\fl.%=-b_{\rm r}\,\fd_{\rm r}\otimes\fl.
\end{equation}
Assuming, for simplicity, the same slip on both slip systems, i.e., \f{\gamma^{\rm A}=\gamma^{\rm B}=\gamma}, gives
\begin{equation}\label{eq:br}
 \f{|\fG|^2}=\gamma^2(\fd^{\rm B}-\fd^{\rm A})\cdot(\fd^{\rm B}-\fd^{\rm A}) \fl\cdot\fl=\gamma^2|\fb_{\rm r}|^2.
\end{equation}
Note that the GB free energy with respect to \f{|\fG|} can, {thus,} be expressed in dependence of~{\f{|\fb_{\rm r}|}}, the magnitude of the RBV~\f{\fb_{\rm r}}, for the special case under consideration. Consequently, Gurtin's GB theory takes into account the residual dislocation content of the GB.\\
The quantity~\f{|\fG|^2} can most generally be expressed by (cf.~\cite{gurtin2008theory})
\begin{equation}
\label{eq:g2}
 |\fG|^2=\sum\limits_{\alpha,\beta}\left(C_{\alpha\beta}^{\rm AA}\gamma_\alpha^{\rm A}\gamma_\beta^{\rm A}+C_{\alpha\beta}^{\rm BB}\gamma^{\rm B}_\alpha\gamma^{\rm B}_\beta-2C_{\alpha\beta}^{\rm AB}\gamma^{\rm A}_\alpha\gamma^{\rm B}_\beta\right),
\end{equation}
which depends on the {\it intra}-action coefficients~\f{C_{\alpha\beta}^{\rm AA},C_{\alpha\beta}^{\rm BB}} and on the {\it inter}-action coefficients~\f{C_{\alpha\beta}^{\rm AB}}. Following from \eqref{eq:g2} and from the discussion below \eqref{eq:CABG}, it can be concluded that Gurtin's theory of GBs has a mechanism to account for the geometric slip transmission criterion~\eqref{eq:Mji}.\\
For the case {under consideration}, the {\it intra}-action coefficients are~\f{C^{\rm AA}=C^{\rm BB}=1} while all other {\it intra}-action coefficients vanish. The {\it inter}-action coefficients vanish, as well, except for $C^{\rm AB}=C^{\rm BA}=\fd^{\rm A}\cdot\fd^{\rm B}$. Equation~\eqref{eq:g2} then reads
\begin{equation} \label{eq:somequation}
|\fG|^2=2\gamma^2\left(1-(\fd^{\rm A}\cdot\fd^{\rm B})\right).
\end{equation}
{Combining \eqref{eq:somequation} with \eqref{eq:br} yields~\f{|\fb_{\rm r}|^2=2\left(1-(\fd^{\rm A}\cdot\fd^{\rm B})\right)}. For the special case of coinciding slip directions, \f{\fd^{\rm A}=\fd^{\rm B}}, this gives~\f{|\fb_{\rm r}|^2=0}, and for the case of perpendicular slip directions, \f{|\fb_{\rm r}|^2=2} is obtained.} Thus, the GB RBV magnitude is a function of the mismatch between slip systems in adjacent grains. The GB Burgers tensor magnitude \f{|\fG|} is a function of the mismatch as well, as is the GB free energy of~\cite{gurtin2008theory} formulated with respect to this quantity.\\
Concluding, it can be stated that Gurtin's GB theory considers the \rthrC{geometrical slip transmission} factor~\eqref{eq:Mji} in the formulation of the GB free energy (via {\it inter}-action coefficients). They are also incorporated in the formulation of the flow rule~\cite{gurtin2008theory}. In addition, the RSSes on the outgoing slip systems (microforce balance / flow rule) are considered in the theory, as is the RBV left in the GB upon a transmission event (GB free energy).
\section{\rthrC{Comparison of geometric criteria for the single slip case}}\label{sec:comparison}
\rthrC{The geometric transmission factors from the preceding sections are compared for a single slip case in order to discuss their differences and limitations in more detail with regard to the crystallographic orientation of the grains and the grain boundary. In the following, it is assumed that slip is occurring on the incoming slip system of grain~A and that, subsequently, slip is activated on the outgoing slip system in grain~B. Four cases are considered as depicted in \figref{fig:rotations}~a)-d). In the cases a)-c), the slip system in grain~B is rotated by the angle~\f{\varphi} about the depicted~\f{\fr}-axis for each case. In a separate case, the grain boundary is rotated by angle~\f{\varphi} about the \f{\fr}-axis depicted in~\figref{fig:rotations}~d). For this case, at first, the slip system in grain~B is left unaltered to isolate the influence of the GB inclination. Then, the slip system in grain~B is pre-rotated by an angle of~\f{15^\circ} about an arbitrarily chosen axis ~\f{\fa=\fe_1+\fe_2+\fe_3}, and the influence of the rotation of the GB is investigated again. By this approach, a more general case than the ideal alignment of the two slip systems is considered.
\begin{figure}[htbp]
\begin{center}{
% \psfrag{n1}{\f{{\color{Emerald}\fn^{\rm A}}}}
% \psfrag{d1}{\f{{\color{Emerald}\fd^{\rm A}}}}
% \psfrag{n2}{\f{{\color{red}\fn^{\rm B}}}}
% \psfrag{d2}{\f{{\color{red}\fd^{\rm B}}}}
% \psfrag{n2t}{\f{{\color{blue}\tilde\fn^{\rm B}}}}
% \psfrag{d2t}{\f{{\color{blue}\tilde\fd^{\rm B}}}}
% \psfrag{ngb}{\f{{\color{black}\fn_\Gamma}}}
% \psfrag{d}{\f{{\color{black}\delta}}}
% \psfrag{e}{\f{{\color{black}\omega}}}
% \psfrag{k}{\f{{\color{black}\kappa}}}
% \psfrag{a}{a)}
% \psfrag{p}{\f{\varphi}}
% \psfrag{b}{b)}
% \psfrag{c}{c)}
% \psfrag{d}{d)}
% \psfrag{e1}{$\fr$}
% \psfrag{e2}{$\fr$}
% \psfrag{e3}{$\fr$}
% \psfrag{x}{$\fe_1$}
% \psfrag{y}{$\fe_2$}
% \psfrag{z}{$\fe_3$}
% \includegraphics[width=0.95\linewidth]{Figure_05.eps}}
\includegraphics[width=0.95\linewidth]{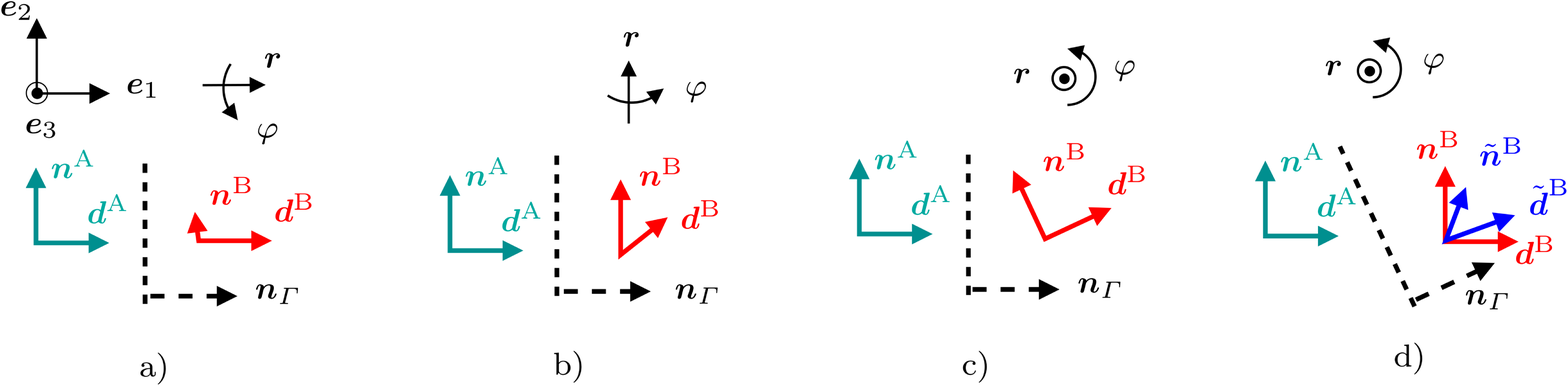}}
\end{center}
\caption{\rthrC{Rotations employed on a single slip system in grain~B and on the grain boundary, respectively, in order to compare the geometric transmission factors. The vectors~\f{\tilde\fn^{\rm B}} and ~\f{\tilde\fd^{\rm B}} are obtained by rotating \f{\fn^{\rm B},\fd^{\rm B}} by $15^\circ$ about the axis~\f{\fa=\fe_1+\fe_2+\fe_3}.}}
\label{fig:rotations}
\end{figure}\\
Plots of the geometric transmission factors are depicted in \figref{fig:plots}. Each column shows results of all geometric factors for the respective cases depicted in \figref{fig:rotations}. The factors are, additionally, referenced by the corresponding equation numbers in \figref{fig:plots}. Since single slip is investigated, the sums of some of the geometric factors contain only one component. Thus, weighting with, e.g., Schmid factors would not appear to be applicable here. In the same spirit, for each geometric factor matrix, only the single occurring component is investigated.}
\begin{figure}
\begin{center}{
\includegraphics[width=0.95\linewidth]{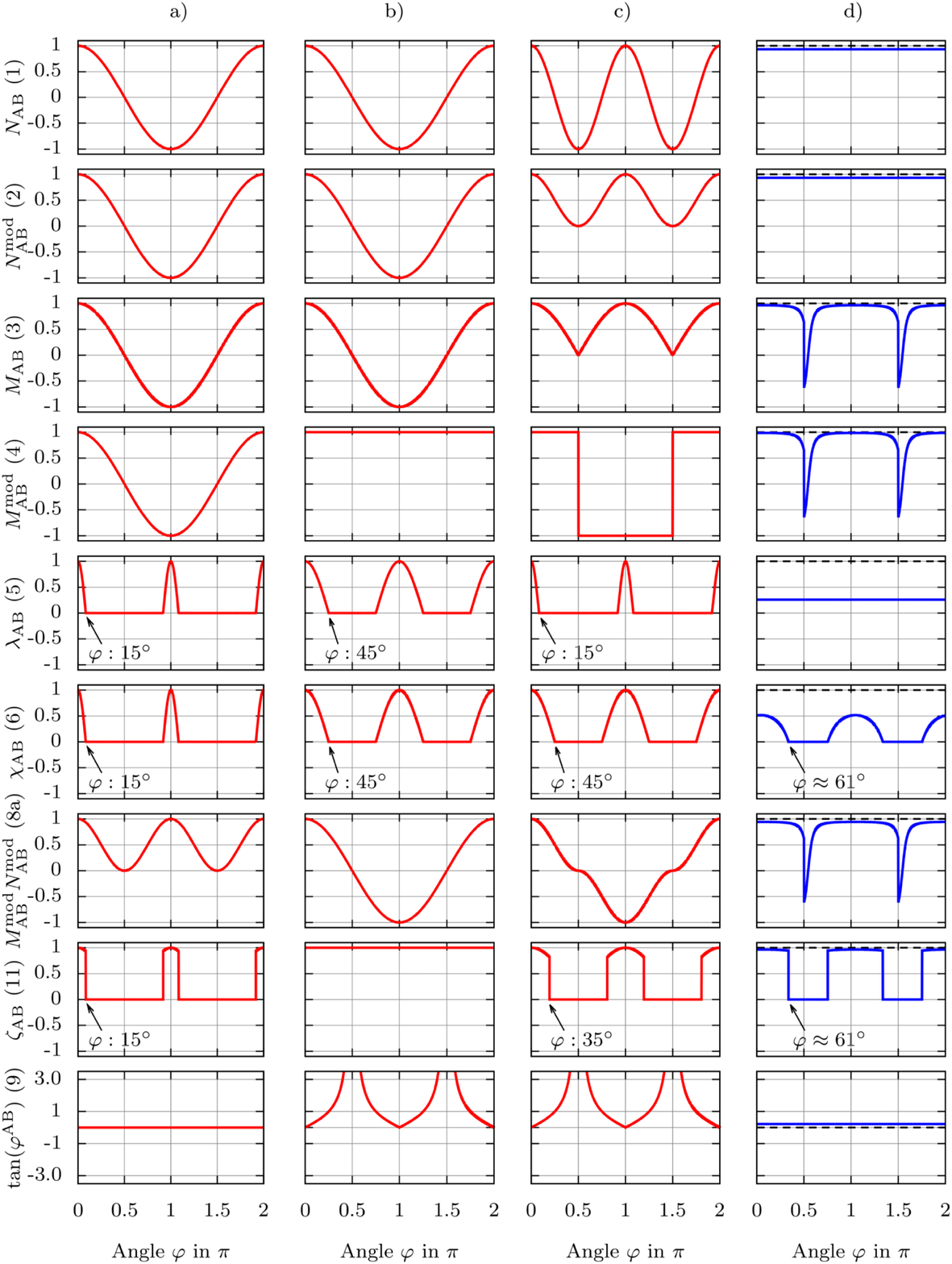}}
\end{center}
        \vspace{-4mm}
\caption{\rthrC{Plots of geometric transmission factors~\eqref{eq:Nji}-\eqref{eq:chi},~\eqref{eq:LRBgamma},~\eqref{eq:tan}, and~\eqref{eq:zeta} for single slip. Rotations employed on slip system of grain~B (first three columns) according to \figref{fig:rotations}~a-c), and on grain boundary (last column) according to \figref{fig:rotations}~d). The dashed lines correspond to a pure rotation of the grain boundary for identical slip systems. The solid lines (in the last column) correspond to a rotation of the grain boundary for the pre-rotated slip system in grain~B, see \figref{fig:rotations}~d).}}
\label{fig:plots}
\end{figure}\\
% \begin{figure}
% \includegraphics[width=\linewidth]{Figure_06_.eps}
% \caption{\rthrC{Plots of geometric transmission factors~\eqref{eq:Nji}-\eqref{eq:chi},~\eqref{eq:LRBgamma},~\eqref{eq:tan}, and~\eqref{eq:zeta} for single slip. Rotations employed on slip system of grain~B (first three columns) according to \figref{fig:rotations}~a-c), and on grain boundary (last column) according to \figref{fig:rotations}~d). The dashed lines correspond to a pure rotation of the grain boundary for identical slip systems. The solid lines (in the last column) correspond to a rotation of the grain boundary for the pre-rotated slip system in grain~B, see \figref{fig:rotations}~d).}}
% \label{fig:plots}
% \end{figure}\\
\rthrC{For a rotation of the slip system~B about the slip direction (\figref{fig:rotations}~a)), the geometric factors~\eqref{eq:Nji}-\eqref{eq:Mjitilde} predict the same behavior. At~\f{\varphi=90^\circ}, all these factors vanish due to the slip plane normals being perpendicular to each other. Factor~\eqref{eq:LRBgamma} also shows essentially the same behavior but its values remain positive due to the multiplication of two (negative) factors. For~\f{\varphi=180^\circ}, all the preceding factors predict ideal alignment, taking into account opposing slip plane normals. The factors~\eqref{eq:lambda},~\eqref{eq:chi} and~\eqref{eq:zeta} also vanish at~\f{\varphi=90^\circ}. The angular region, however, where slip transmission can occur, is far less pronounced than for the previous factors. This is due to the inherent limits, imposed by the critical angles for slip transmission, see, e.g.,~\eqref{eq:lambda} in \figref{fig:plots}, column~a). In addition, the factors~\eqref{eq:lambda},~\eqref{eq:chi}, and~\eqref{eq:zeta} are different in their transition behavior: while~\eqref{eq:lambda},~\eqref{eq:chi} predict a steep but smooth transition from possible slip transfer to no slip transfer,~\eqref{eq:zeta} shows a less steep transition at first, then a jump to ``no transmission``. This is caused by the scaling that is included in~\eqref{eq:lambda},~\eqref{eq:chi} but not in~\eqref{eq:zeta}. The factor~\eqref{eq:tan} vanishes for this case. This corresponds to fully possible transmission since~\eqref{eq:tan} is used in the resistance of the grain boundary against slip transmission. However, the applicability of this factor is limited due to the neglect of the slip plane orientations.\\
Rotating the slip direction by angle~\f{\varphi} about the slip plane normal (\figref{fig:rotations}~b)) gives identical results for factors~\eqref{eq:Nji}-\eqref{eq:Mji} as in the previous case. However,~\eqref{eq:Mjitilde} is not affected by this rotation since it only considers the lines of intersection with the GB, and not the slip directions. One should, nevertheless, keep in mind that~\eqref{eq:Mjitilde} is combined with additional considerations regarding the slip directions (see \figref{fig:slipcriteria}). The product~\eqref{eq:LRBgamma} of~\eqref{eq:Mjitilde} and~\eqref{eq:Njitilde} is identical to~\eqref{eq:Mjitilde}, as a consequence of~\eqref{eq:Mjitilde} being constant for the investigated case. The factors~\eqref{eq:lambda},~\eqref{eq:chi} again show a transition behavior that is, however, less steep due to the higher critical angle employed for slip directions. Contrary to the case of rotation~a),~\eqref{eq:zeta} is not affected by the rotation and is, therefore, constant. This appears to be unphysical and is caused by the neglect of the slip directions in this factor. For the employed rotation of the slip direction,~\eqref{eq:tan} approaches infinite resistance of the GB against slip transmission at~\f{\varphi=90^\circ}.\\
For a rotation of both slip direction and slip plane normal about an axis perpendicular to them (\figref{fig:rotations}~c)),~\eqref{eq:Nji} predicts optimal transmission at \f{\varphi=90^\circ}. This appears to be unphysical since both slip direction and slip plane normal in grain~B are perpendicular to their counterparts in grain~A which would severely restrict transmission. The factors~\eqref{eq:Njitilde}-\eqref{eq:Mji}, however, predict no transmission at this angle. Factor~\eqref{eq:Mjitilde} is constant for this case since the lines of intersection do not change for the employed rotation. The depicted jump is of numerical nature and does not alter the qualitative behavior. Factor~\eqref{eq:lambda} considers two critical angles, and thus, in this case the smaller angle limits the transmission regime. Although~\eqref{eq:chi} also takes into account two critical angles, the larger one limits the transmission in the investigated case. This is due to the consideration of the angle between the lines of intersection (which do not change in this case) and the neglect of slip directions. The product~\eqref{eq:LRBgamma} shows a slightly larger regime of no transmission, compared to the previous two criteria. For~\eqref{eq:zeta}, the critical angle to be considered is the angle between the slip directions since the lines of intersection are unaltered. This gives a slightly larger regime of possible slip transmission than for rotation~a). In this context, it is also questionable why limit values should be employed for both the lines of intersection and the slip plane normals. The angle between slip plane normals is always larger than the angle between the lines of intersection (see also \cite{werner1990slip}). The criterion~\eqref{eq:tan} shows identical behavior as in the previous case. This is also questionable since the alignment of slip planes has been found to be more substantial in predicting slip transmission and, thus, would be expected to limit transmissibility upon applying identical rotation to the slip direction and the slip plane normal.\\
In the last case, the grain boundary is rotated about an axis perpendicular to the slip directions and the slip plane normals of both grains (\figref{fig:rotations}~d)). The slip systems, however, are not altered, at first. This case, thus, isolates the influence of the GB when comparing the different geometrical transmission factors. Interestingly, the orientation of the GB does not influence any of the investigated factors for this case (dashed lines in~\figref{fig:plots}, column~d)).\\
Then, the slip system in grain~B is pre-rotated by~\f{15^\circ} about an axis~\f{\fa} (\figref{fig:rotations}~d)). The GB is rotated again about the depicted axis~\f{\fr} (\figref{fig:rotations}~d)). Both factors~\eqref{eq:Nji} and~\eqref{eq:Njitilde} are not affected by this rotation since they do not account for the GB orientation. Both give constant results for this case (slightly smaller than ''one``, since the slip directions are only changed initially). Factor~\eqref{eq:lambda} does also not account for the GB orientation and shows a constant result. Factors~\eqref{eq:Mji} and~\eqref{eq:Mjitilde}, however, both account for the GB orientation. The product~\eqref{eq:LRBgamma} also considers the orientation of the GB. Factor~\eqref{eq:chi} is clearly affected by the GB orientation, and factor~\eqref{eq:zeta} also shows slight changes upon the employed rotation of the GB. An effective cut-off angle of~\f{\varphi\approx61^\circ} is obtained for both~\eqref{eq:chi} and~\eqref{eq:zeta} (see \figref{fig:plots}~d)) due to the inherent limiting angle for the lines of intersection alignment. Furthermore, the GB orientation is not considered by~\eqref{eq:tan} at all. The slightly higher value than in the previous case is a result of the employed initial rotation for the slip direction in grain~B.
}

%%%% REFERENCES TO FACTORS
%~\eqref{eq:Nji}		(1)
%~\eqref{eq:Njitilde}		(2)
%~\eqref{eq:Mji}		(3)
%~\eqref{eq:Mjitilde}		(4)
%~\eqref{eq:lambda}		(5)
%~\eqref{eq:chi}		(6)
%~\eqref{eq:LRBgamma}		(8a)
%~\eqref{eq:zeta}		(11)
%~\eqref{eq:tan}		(9)

\vspace{-2mm}
\section{Conclusion}
In the past, geometric criteria have been shown to be the most important ones in predicting the slip transmission across grain boundaries in experiments \jnew{on metals} \revC{as well as computational investigations using discrete methods}. {Here, a comprehensive overview on \rthrC{the \revC{geometric} criteria used in both experiments} and computational models, \revC{focusing on continuum approaches}, is given in the work at hand.}\\
\rthrC{Regarding the experimental and computational verification of slip transmission criteria, there are more investigations dealing with specific grain boundary types, e.g., symmetrical boundaries~\cite{koning2002modelling,jin2008interactions,cheng2008atomistic}. Although some experimental works~\cite{gibson2002slip,soer2005incipient} and recent computational works~\cite{brandl2007slip,bachurin2010dislocation} also consider more general grain boundaries, various findings exist in regard to the influence of the grain boundary type. While in~\cite{lim1985continuity,gemperlova2004slip,gemperle2005reactions} transmission was found to depend on the type of the investigated symmetric grain boundaries, in~\cite{wo2004investigation}, the behavior of coincidence site lattice boundaries and general boundaries did not appear to be substantially different regarding slip transmission.}\\
\rthrC{Open questions from an experimental point of view also include the evaluation of continuity of slip traces. Several types of continuity and discontinuity exist, as briefly discussed in the work at hand and shown in detail, e.g., in~\cite{west2013strain}. Computational models that incorporate slip transmission, consequently, also need to define ad hoc the continuity of plastic slip.}\\
\rthrC{In the context of embedding slip transmission criteria in continuum models, several approaches exist. The \bnew{detailed} comparison of Gurtin's grain boundary model~\cite{gurtin2008theory} \bnew{to} \rthrC{the slip transmission} criteria \bnew{shows} that this theory can be interrelated to the three main criteria \rthrC{used in experiments}. Common to most continuum models is the incorporation of the resolved shear stresses due to the inherent modeling of plastic slip with flow rules. This can be combined with geometric transmission factors and, e.g., as in the case of~\cite{gurtin2008theory}, a grain boundary energy considering the residual Burgers vector. Open questions for continuum models include the appropriate balance between computational effort and the necessary level of physical enrichment of interface kinematics. The latter can be obtained from consideration of experiments and atomistic simulations showing the discrete mechanisms of slip transfer. The broad basis of experimental and computational investigations in regard to~\eqref{eq:Mji},~\eqref{eq:Mjitilde}, and the natural incorporation of it in~\cite{gurtin2008theory} in combination with the other criteria outlined recommend further investigation of this approach in extended simulations.}\\
\rthrC{The comparison of the geometric criteria in this work results in various findings. First of all, the geometric criterion~\eqref{eq:Nji} contradicts the common understanding of slip transmission since it predicts fully possible transmission for two slip systems with slip directions being perpendicular to each other. The factor~\eqref{eq:Njitilde}, however, predicts no transmission for this case but neglects the orientation of the grain boundary. Factors~\eqref{eq:Mji} and~\eqref{eq:Mjitilde} account for the orientation of the grain boundary and have shown better agreement with experiments if combined further with additional considerations regarding the residual Burgers vector and the resolved shear stresses~\cite{lee1989prediction,lee1990tem,clark1992criteria}. Whether the inclination of the grain boundary needs to be considered in such criteria is debatable. In~\cite{bachurin2010dislocation}, for example, no influence of the grain boundary inclination on the slip transmission process was found. Indications in this direction are also present in~\cite{liu1995dislocation}, where the influence of the grain boundary orientation with regard to the loading direction was found to be of minor importance for the nucleation processes investigated in ice crystals. A further open question is whether critical angles need to be used, e.g., employed in~\eqref{eq:lambda},~\eqref{eq:chi}, and~\eqref{eq:zeta}. These criteria predict a far less pronounced regime for possible slip transmission than criteria without such angles. In addition, the experimental foundation for the critical angles is not exhaustive. In fact, some of them (e.g.,~\cite{werner1990slip}) have been determined based on an observed range of possible slip transmission~\cite{davis1966slip}. Criterion~\eqref{eq:zeta} does not seem to be applicable for general slip transmission processes since the orientation of the slip directions is not considered which contradicts experiments. In the same spirit,~\eqref{eq:tan} neglects the slip plane orientations and thus is not applicable when the slip planes for a transmission process are distinct. The use of (weighted) sums of geometric slip transmission factors necessitates further investigations. In~\cite{bieler2014grain}, similar distributions of such weighted approaches were found, although different geometric factors and weights were employed. Furthermore, there are apparent contradictions between the use of~\eqref{eq:lambda} and experimental findings~\cite{kumar2010influence}, also questioning the use of this factor. It would be desirable to compare the geometric transmission criteria including additional criteria (RSS, RBV) for multiple slip systems, preferably also for different crystal structures.
}\\[3ex]

\begin{acknowledgements}
% \newline
The authors acknowledge the following support:\\
T.~B\"ohlke and E.~Bayerschen: German Research Foundation~(DFG) under Grant BO1466/5-1, part of the DFG Research Group 1650 "Dislocation based Plasticity", and Karlsruhe House of Young Scientists (KHYS) for funding a stay of the first author of this work at CERECAM, UCT. \revC{The support by E.~Ramani in preparing the plots in \secref{sec:comparison} is acknowledged.}\\
B.D.~Reddy and A.T.~McBride: National Research Foundation of South Africa (SA) through the SA Research Chair in Computational Mechanics.\\
\end{acknowledgements}

\bibliographystyle{myunsrt}
\bibliography{lit_full}   % name your BibTeX data base

\begin{thebibliography}{100}

\bibitem{hirth1972influence}
J.~P. Hirth.
\newblock The influence of grain boundaries on mechanical properties.
\newblock {\em Metallurgical Transactions}, 3(12):3047--3067, 1972.

\bibitem{shen1988dislocation}
Z.~Shen, R.~Wagoner, and W.~Clark.
\newblock Dislocation and grain boundary interactions in metals.
\newblock {\em Acta Metallurgica}, 36(12):3231--3242, 1988.

\bibitem{davis1966slip}
K.~Davis, E.~Teghtsoonian, and A.~Lu.
\newblock Slip band continuity across grain boundaries in aluminum.
\newblock {\em Acta Metallurgica}, 14(12):1677--1684, 1966.

\bibitem{lim1985interaction}
L.~Lim and R.~Raj.
\newblock Interaction between lattice and grain boundary dislocations and their
  role in mechanical properties of interfaces.
\newblock {\em Le Journal de Physique Colloques}, 46(C4):C4--581, 1985.

\bibitem{shen1986dislocation}
Z.~Shen, R.~Wagoner, and W.~Clark.
\newblock Dislocation pile-up and grain boundary interactions in 304 stainless
  steel.
\newblock {\em Scripta Metallurgica}, 20(6):921--926, 1986.

\bibitem{murr1981strain}
L.~Murr.
\newblock Strain-induced dislocation emission from grain boundaries in
  stainless steel.
\newblock {\em Materials Science and Engineering}, 51(1):71--79, 1981.

\bibitem{lim1984slip}
L.~Lim.
\newblock Slip-twin interactions in nickel at 573{K} at large strains.
\newblock {\em Scripta Metallurgica}, 18(10):1139--1142, 1984.

\bibitem{lee1989anomalous}
T.~Lee, I.~Robertson, and H.~Birnbaum.
\newblock Anomalous slip in an {FCC} system.
\newblock {\em Ultramicroscopy}, 29(1):212--216, 1989.

\bibitem{medlin1997climb}
D.~Medlin, C.~Carter, J.~Angelo, and M.~Mills.
\newblock Climb and glide of a/3\textless111\textgreater dislocations in an
  aluminium {{$\Sigma$}}= 3 boundary.
\newblock {\em Philosophical Magazine A}, 75(3):733--747, 1997.

\bibitem{zghal2001transmission}
S.~Zghal, A.~Coujou, and A.~Couret.
\newblock Transmission of the deformation through $\gamma$-$\gamma$ interfaces
  in a polysynthetically twinned {TiAl} alloy {I. O}rdered domain interfaces
  (120$^\circ$ rotational).
\newblock {\em Philosophical Magazine A}, 81(2):345--364, 2001.

\bibitem{zghal2001transmissionb}
S.~Zghal and A.~Couret.
\newblock Transmission of the deformation through $\gamma$-$\gamma$ interfaces
  in a polysynthetically twinned {TiAl} alloy {II. T}win interfaces
  (180$^\circ$ rotational).
\newblock {\em Philosophical Magazine A}, 81(2):365--382, 2001.

\bibitem{lim1985continuity}
L.~Lim and R.~Raj.
\newblock Continuity of slip screw and mixed crystal dislocations across
  bicrystals of nickel at 573 {K}.
\newblock {\em Acta Metallurgica}, 33(8):1577--1583, 1985.

\bibitem{gemperlova2004slip}
J.~Gemperlova, M.~Polcarova, A.~Gemperle, and N.~Zarubova.
\newblock Slip transfer across grain boundaries in fe--si bicrystals.
\newblock {\em Journal of Alloys and Compounds}, 378(1):97--101, 2004.

\bibitem{gemperle2005reactions}
A.~Gemperle, N.~Zarubova, and J.~Gemperlova.
\newblock Reactions of slip dislocations with twin boundary in {Fe-Si}
  bicrystals.
\newblock {\em Journal of Materials Science}, 40(12):3247--3254, 2005.

\bibitem{pond2006study}
R.~C. Pond, D.~L. Medlin, and A.~Serra.
\newblock A study of the accommodation of coherency strain by interfacial
  defects at a grain boundary in gold.
\newblock {\em Philosophical Magazine}, 86(29-31):4667--4684, 2006.

\bibitem{takasugi1978activated}
T.~Takasugi, O.~Izumi, and N.~Fat-Halla.
\newblock Activated slip systems during yielding of $\alpha$-$\beta$ brass
  two-phase bicrystals.
\newblock {\em Journal of Materials Science}, 13(9):2013--2021, 1978.

\bibitem{forwood1981prismatic}
C.~Forwood and L.~Clarebrough.
\newblock Prismatic glide and slip transfer across a high-angle grain boundary.
\newblock {\em Philosophical Magazine A}, 44(1):31--41, 1981.

\bibitem{de2006situ}
J.~T. De~Hosson, W.~A. Soer, A.~M. Minor, Z.~Shan, E.~A. Stach, S.~S. Asif, and
  O.~L. Warren.
\newblock In situ {TEM} nanoindentation and dislocation-grain boundary
  interactions: a tribute to {D}avid {B}randon.
\newblock {\em Journal of Materials Science}, 41(23):7704--7719, 2006.

\bibitem{lim1985role}
L.~Lim and R.~Raj.
\newblock The role of residual dislocation arrays in slip induced cavitation,
  migration and dynamic recrystallization at grain boundaries.
\newblock {\em Acta Metallurgica}, 33(12):2205--2214, 1985.

\bibitem{pond1977absorption}
R.~Pond and D.~Smith.
\newblock On the absorption of dislocations by grain boundaries.
\newblock {\em Philosophical Magazine}, 36(2):353--366, 1977.

\bibitem{clark1979interaction}
W.~Clark and D.~Smith.
\newblock Interaction of lattice dislocations with periodic grain boundary
  structures.
\newblock {\em Journal of Materials Science}, 14(4):776--788, 1979.

\bibitem{malis1979grain}
T.~Malis and K.~Tangri.
\newblock Grain boundaries as dislocation sources in the premacroyield strain
  region.
\newblock {\em Acta Metallurgica}, 27(1):25--32, 1979.

\bibitem{brentnall1965some}
W.~Brentnall and W.~Rostoker.
\newblock Some observations on microyielding.
\newblock {\em Acta Metallurgica}, 13(3):187--198, 1965.

\bibitem{lall1979orientation}
C.~Lall, S.~Chin, and D.~Pope.
\newblock The orientation and temperature dependence of the yield stress of
  {Ni3} {(Al, Nb)} single crystals.
\newblock {\em Metallurgical Transactions A}, 10(9):1323--1332, 1979.

\bibitem{seal2012analysis}
J.~R. Seal, M.~A. Crimp, T.~R. Bieler, and C.~J. Boehlert.
\newblock Analysis of slip transfer and deformation behavior across the
  $\alpha$/$\beta$ interface in {T}i--5{A}l--2.5 {S}n (wt.\%) with an equiaxed
  microstructure.
\newblock {\em Materials Science and Engineering: A}, 552:61--68, 2012.

\bibitem{west2013strain}
E.~West and G.~Was.
\newblock Strain incompatibilities and their role in intergranular cracking of
  irradiated {316L} stainless steel.
\newblock {\em Journal of Nuclear Materials}, 441(1):623--632, 2013.

\bibitem{bridier2005analysis}
F.~Bridier, P.~Villechaise, and J.~Mendez.
\newblock Analysis of the different slip systems activated by tension in a
  $\alpha$/$\beta$ titanium alloy in relation with local crystallographic
  orientation.
\newblock {\em Acta Materialia}, 53(3):555--567, 2005.

\bibitem{abuzaid2012slip}
W.~Z. Abuzaid, M.~D. Sangid, J.~D. Carroll, H.~Sehitoglu, and J.~Lambros.
\newblock Slip transfer and plastic strain accumulation across grain boundaries
  in {H}astelloy {X}.
\newblock {\em Journal of the Mechanics and Physics of Solids},
  60(6):1201--1220, 2012.

\bibitem{miura1978plastic}
S.~Miura and Y.~Saeki.
\newblock Plastic deformation of aluminum bicrystals \textless100\textgreater
  oriented.
\newblock {\em Acta Metallurgica}, 26(1):93--101, 1978.

\bibitem{clark1992criteria}
W.~Clark, R.~Wagoner, Z.~Shen, T.~Lee, I.~Robertson, and H.~Birnbaum.
\newblock On the criteria for slip transmission across interfaces in
  polycrystals.
\newblock {\em Scripta Metallurgica et Materialia}, 26(2):203--206, 1992.

\bibitem{guo2014slip}
Y.~Guo, T.~Britton, and A.~Wilkinson.
\newblock Slip band--grain boundary interactions in commercial-purity titanium.
\newblock {\em Acta Materialia}, 76:1--12, 2014.

\bibitem{lee1989prediction}
T.~Lee, I.~Robertson, and H.~Birnbaum.
\newblock Prediction of slip transfer mechanisms across grain boundaries.
\newblock {\em Scripta Metallurgica}, 23(5):799--803, 1989.

\bibitem{kacher2012quasi}
J.~Kacher and I.~Robertson.
\newblock Quasi-four-dimensional analysis of dislocation interactions with
  grain boundaries in 304 stainless steel.
\newblock {\em Acta Materialia}, 60(19):6657--6672, 2012.

\bibitem{zhou2012dislocation}
C.~Zhou and R.~LeSar.
\newblock Dislocation dynamics simulations of plasticity in polycrystalline
  thin films.
\newblock {\em International Journal of Plasticity}, 30:185--201, 2012.

\bibitem{bachurin2010dislocation}
D.~Bachurin, D.~Weygand, and P.~Gumbsch.
\newblock Dislocation--grain boundary interaction in \textless111\textgreater
  textured thin metal films.
\newblock {\em Acta Materialia}, 58(16):5232--5241, 2010.

\bibitem{sangid2011energy}
M.~D. Sangid, T.~Ezaz, H.~Sehitoglu, and I.~M. Robertson.
\newblock Energy of slip transmission and nucleation at grain boundaries.
\newblock {\em Acta Materialia}, 59(1):283--296, 2011.

\bibitem{koning2002modelling}
M.~d. Koning, R.~Miller, V.~Bulatov, and F.~F. Abraham.
\newblock Modelling grain-boundary resistance in intergranular dislocation slip
  transmission.
\newblock {\em Philosophical Magazine A}, 82(13):2511--2527, 2002.

\bibitem{brandl2007slip}
C.~Brandl, E.~Bitzek, P.~Derlet, and H.~Van~Swygenhoven.
\newblock Slip transfer through a general high angle grain boundary in
  nanocrystalline aluminum.
\newblock {\em Applied Physics Letters}, 91(11):111914, 2007.

\bibitem{zhu2012plastic}
T.~Zhu and H.~Gao.
\newblock Plastic deformation mechanism in nanotwinned metals: an insight from
  molecular dynamics and mechanistic modeling.
\newblock {\em Scripta Materialia}, 66(11):843--848, 2012.

\bibitem{baillin1987dislocation}
X.~Baillin, J.~Pelissier, J.~Bacmann, A.~Jacques, and A.~George.
\newblock Dislocation transmission through {$\Sigma$}= 9 symmetrical tilt
  boundaries in silicon and germanium: {I}. {I}n situ observations by
  synchrotron {X}-ray topography and high-voltage electron microscopy.
\newblock {\em Philosophical Magazine A}, 55(2):143--164, 1987.

\bibitem{couzinie2005interaction}
J.~Couzinie, B.~Decamps, and L.~Priester.
\newblock Interaction of dissociated lattice dislocations with a {$\Sigma$}= 3
  grain boundary in copper.
\newblock {\em International Journal of Plasticity}, 21(4):759--775, 2005.

\bibitem{jin2006interaction}
Z.-H. Jin, P.~Gumbsch, E.~Ma, K.~Albe, K.~Lu, H.~Hahn, and H.~Gleiter.
\newblock The interaction mechanism of screw dislocations with coherent twin
  boundaries in different face-centred cubic metals.
\newblock {\em Scripta Materialia}, 54(6):1163--1168, 2006.

\bibitem{dewald2007multiscale}
M.~Dewald and W.~Curtin.
\newblock Multiscale modelling of dislocation/grain-boundary interactions: I.
  edge dislocations impinging on {$\Sigma$}11 (1 1 3) tilt boundary in {Al}.
\newblock {\em Modelling and Simulation in Materials Science and Engineering},
  15(1):S193, 2007.

\bibitem{dewald2007multiscaleb}
M.~Dewald and W.~Curtin.
\newblock Multiscale modelling of dislocation/grain boundary interactions. {II.
  S}crew dislocations impinging on tilt boundaries in {A}l.
\newblock {\em Philosophical Magazine}, 87(30):4615--4641, 2007.

\bibitem{liu2012simulation}
B.~Liu, P.~Eisenlohr, F.~Roters, and D.~Raabe.
\newblock Simulation of dislocation penetration through a general low-angle
  grain boundary.
\newblock {\em Acta Materialia}, 60(13):5380--5390, 2012.

\bibitem{kacher2014situ}
J.~Kacher and I.~M. Robertson.
\newblock In situ and tomographic analysis of dislocation/grain boundary
  interactions in $\alpha$-titanium.
\newblock {\em Philosophical Magazine}, 94(8):814--829, 2014.

\bibitem{soer2005detection}
W.~Soer and J.~T.~M. De~Hosson.
\newblock Detection of grain-boundary resistance to slip transfer using
  nanoindentation.
\newblock {\em Materials Letters}, 59(24):3192--3195, 2005.

\bibitem{britton2009nanoindentation}
T.~Britton, D.~Randman, and A.~Wilkinson.
\newblock Nanoindentation study of slip transfer phenomenon at grain
  boundaries.
\newblock {\em Journal of Materials Research}, 24(03):607--615, 2009.

\bibitem{kacher2014dislocation}
J.~Kacher, B.~Eftink, B.~Cui, and I.~Robertson.
\newblock Dislocation interactions with grain boundaries.
\newblock {\em Current Opinion in Solid State and Materials Science},
  18(4):227--243, 2014.

\bibitem{misra1999slip}
A.~Misra and R.~Gibala.
\newblock Slip transfer and dislocation nucleation processes in multiphase
  ordered {Ni-Fe-Al} alloys.
\newblock {\em Metallurgical and Materials Transactions A}, 30(4):991--1001,
  1999.

\bibitem{beyerlein2013mapping}
I.~J. Beyerlein, J.~Wang, and R.~Zhang.
\newblock Mapping dislocation nucleation behavior from bimetal interfaces.
\newblock {\em Acta Materialia}, 61(19):7488--7499, 2013.

\bibitem{beyerlein2014influence}
I.~Beyerlein, J.~Mayeur, R.~McCabe, S.~Zheng, J.~Carpenter, and N.~Mara.
\newblock Influence of slip and twinning on the crystallographic stability of
  bimetal interfaces in nanocomposites under deformation.
\newblock {\em Acta Materialia}, 72:137--147, 2014.

\bibitem{beyerlein2012structure}
I.~Beyerlein, N.~Mara, J.~Wang, J.~Carpenter, S.~Zheng, W.~Han, R.~Zhang,
  K.~Kang, T.~Nizolek, and T.~Pollock.
\newblock Structure--property--functionality of bimetal interfaces.
\newblock {\em JOM}, 64(10):1192--1207, 2012.

\bibitem{misra2005length}
A.~Misra, J.~Hirth, and R.~Hoagland.
\newblock Length-scale-dependent deformation mechanisms in incoherent metallic
  multilayered composites.
\newblock {\em Acta Materialia}, 53(18):4817--4824, 2005.

\bibitem{asaro1983crystal}
R.~J. Asaro.
\newblock Crystal plasticity.
\newblock {\em Journal of Applied Mechanics}, 50(4b):921--934, 1983.

\bibitem{needleman1993comparison}
A.~Needleman and V.~Tvergaard.
\newblock Comparison of crystal plasticity and isotropic hardening predictions
  for metal-matrix composites.
\newblock {\em Journal of Applied Mechanics}, 60(1):70--76, 1993.

\bibitem{yao2014plastic}
W.~Yao, C.~Krill~III, B.~Albinski, H.-C. Schneider, and J.~You.
\newblock Plastic material parameters and plastic anisotropy of tungsten single
  crystal: a spherical micro-indentation study.
\newblock {\em Journal of Materials Science}, 49(10):3705--3715, 2014.

\bibitem{ziemann2015deformation}
M.~Ziemann, Y.~Chen, O.~Kraft, E.~Bayerschen, S.~Wulfinghoff, C.~Kirchlechner,
  N.~Tamura, T.~B{\"o}hlke, M.~Walter, and P.~Gruber.
\newblock Deformation patterns in cross-sections of twisted bamboo-structured
  {Au} microwires.
\newblock {\em Acta Materialia}, 97:216--222, 2015.

\bibitem{beyerlein2007plastic}
I.~J. Beyerlein, D.~J. Alexander, and C.~N. Tom{\'e}.
\newblock Plastic anisotropy in aluminum and copper pre-strained by equal
  channel angular extrusion.
\newblock {\em Journal of Materials Science}, 42(5):1733--1750, 2007.

\bibitem{eyckens2015prediction}
P.~Eyckens, H.~Mulder, J.~Gawad, H.~Vegter, D.~Roose, T.~van~den Boogaard,
  A.~Van~Bael, and P.~Van~Houtte.
\newblock The prediction of differential hardening behaviour of steels by
  multi-scale crystal plasticity modelling.
\newblock {\em International Journal of Plasticity}, 2015.

\bibitem{kalidindi1998incorporation}
S.~R. Kalidindi.
\newblock Incorporation of deformation twinning in crystal plasticity models.
\newblock {\em Journal of the Mechanics and Physics of Solids}, 46(2):267--290,
  1998.

\bibitem{forest1998modeling}
S.~Forest.
\newblock Modeling slip, kink and shear banding in classical and generalized
  single crystal plasticity.
\newblock {\em Acta Materialia}, 46(9):3265--3281, 1998.

\bibitem{zaafarani2006three}
N.~Zaafarani, D.~Raabe, R.~Singh, F.~Roters, and S.~Zaefferer.
\newblock Three-dimensional investigation of the texture and microstructure
  below a nanoindent in a cu single crystal using {3D EBSD} and crystal
  plasticity finite element simulations.
\newblock {\em Acta Materialia}, 54(7):1863--1876, 2006.

\bibitem{zhang2015multi}
K.~Zhang, B.~Holmedal, O.~S. Hopperstad, S.~Dumoulin, J.~Gawad, A.~Van~Bael,
  and P.~Van~Houtte.
\newblock Multi-level modelling of mechanical anisotropy of commercial pure
  aluminium plate: crystal plasticity models, advanced yield functions and
  parameter identification.
\newblock {\em International Journal of Plasticity}, 66:3--30, 2015.

\bibitem{tian2014dislocation}
L.~Tian, A.~Russell, and I.~Anderson.
\newblock A dislocation-based, strain--gradient--plasticity strengthening model
  for deformation processed metal--metal composites.
\newblock {\em Journal of Materials Science}, 49(7):2787--2794, 2014.

\bibitem{mayeur2015incorporating}
J.~Mayeur, I.~Beyerlein, C.~Bronkhorst, and H.~Mourad.
\newblock Incorporating interface affected zones into crystal plasticity.
\newblock {\em International Journal of Plasticity}, 65:206--225, 2015.

\bibitem{gurtin2008theory}
M.~E. Gurtin.
\newblock A theory of grain boundaries that accounts automatically for grain
  misorientation and grain-boundary orientation.
\newblock {\em Journal of the Mechanics and Physics of Solids}, 56(2):640--662,
  2008.

\bibitem{wulfinghoff2013gradient}
S.~Wulfinghoff, E.~Bayerschen, and T.~B{\"o}hlke.
\newblock A gradient plasticity grain boundary yield theory.
\newblock {\em International Journal of Plasticity}, 51:33--46, 2013.

\bibitem{van2013grain}
P.~Van~Beers, G.~McShane, V.~Kouznetsova, and M.~Geers.
\newblock Grain boundary interface mechanics in strain gradient crystal
  plasticity.
\newblock {\em Journal of the Mechanics and Physics of Solids},
  61(12):2659--2679, 2013.

\bibitem{wolf1990correlation}
D.~Wolf.
\newblock Correlation between structure, energy, and ideal cleavage fracture
  for symmetrical grain boundaries in fcc metals.
\newblock {\em Journal of Materials Research}, 5(08):1708--1730, 1990.

\bibitem{fleck1994strain}
N.~Fleck, G.~Muller, M.~Ashby, and J.~Hutchinson.
\newblock Strain gradient plasticity: theory and experiment.
\newblock {\em Acta Metallurgica et Materialia}, 42(2):475--487, 1994.

\bibitem{chen2015size}
Y.~Chen, O.~Kraft, and M.~Walter.
\newblock Size effects in thin coarse-grained gold microwires under tensile and
  torsional loading.
\newblock {\em Acta Materialia}, 87:78--85, 2015.

\bibitem{gurtin2000plasticity}
M.~E. Gurtin.
\newblock On the plasticity of single crystals: free energy, microforces,
  plastic-strain gradients.
\newblock {\em Journal of the Mechanics and Physics of Solids},
  48(5):989--1036, 2000.

\bibitem{gurtin2002gradient}
M.~E. Gurtin.
\newblock A gradient theory of single-crystal viscoplasticity that accounts for
  geometrically necessary dislocations.
\newblock {\em Journal of the Mechanics and Physics of Solids}, 50(1):5--32,
  2002.

\bibitem{rieger2015microstructure}
F.~Rieger and T.~B{\"o}hlke.
\newblock Microstructure based prediction and homogenization of the strain
  hardening behavior of dual-phase steel.
\newblock {\em Archive of Applied Mechanics}, pages 1--20, 2015.

\bibitem{aifantis2006interfaces}
K.~Aifantis, W.~Soer, J.~T.~M. De~Hosson, and J.~Willis.
\newblock Interfaces within strain gradient plasticity: theory and experiments.
\newblock {\em Acta Materialia}, 54(19):5077--5085, 2006.

\bibitem{zhang2014internal}
X.~Zhang, K.~E. Aifantis, J.~Senger, D.~Weygand, and M.~Zaiser.
\newblock Internal length scale and grain boundary yield strength in gradient
  models of polycrystal plasticity: How do they relate to the dislocation
  microstructure?
\newblock {\em Journal of Materials Research}, 29(18):2116--2128, 2014.

\bibitem{geers2014coupled}
M.~Geers, M.~Cottura, B.~Appolaire, E.~P. Busso, S.~Forest, and A.~Villani.
\newblock Coupled glide-climb diffusion-enhanced crystal plasticity.
\newblock {\em Journal of the Mechanics and Physics of Solids}, 70:136--153,
  2014.

\bibitem{van2015defect}
P.~van Beers, V.~Kouznetsova, and M.~Geers.
\newblock Defect redistribution within a continuum grain boundary plasticity
  model.
\newblock {\em Journal of the Mechanics and Physics of Solids}, 2015.

\bibitem{reuber2014dislocation}
C.~Reuber, P.~Eisenlohr, F.~Roters, and D.~Raabe.
\newblock Dislocation density distribution around an indent in
  single-crystalline nickel: {C}omparing nonlocal crystal plasticity
  finite-element predictions with experiments.
\newblock {\em Acta Materialia}, 71:333--348, 2014.

\bibitem{dogge2015interface}
M.~Dogge, R.~Peerlings, and M.~Geers.
\newblock Interface modeling in continuum dislocation transport.
\newblock {\em Mechanics of Materials}, 88:30--43, 2015.

\bibitem{lucadamo2002dislocation}
G.~Lucadamo and D.~Medlin.
\newblock Dislocation emission at junctions between {$\Sigma$}= 3 grain
  boundaries in gold thin films.
\newblock {\em Acta Materialia}, 50(11):3045--3055, 2002.

\bibitem{wulfinghoff2015gradient}
S.~Wulfinghoff and T.~B{\"o}hlke.
\newblock Gradient crystal plasticity including dislocation-based
  work-hardening and dislocation transport.
\newblock {\em International Journal of Plasticity}, 69:152--169, 2015.

\bibitem{hochrainer2014continuum}
T.~Hochrainer, S.~Sandfeld, M.~Zaiser, and P.~Gumbsch.
\newblock Continuum dislocation dynamics: towards a physical theory of crystal
  plasticity.
\newblock {\em Journal of the Mechanics and Physics of Solids}, 63:167--178,
  2014.

\bibitem{evers2004scale}
L.~Evers, W.~Brekelmans, and M.~Geers.
\newblock Scale dependent crystal plasticity framework with dislocation density
  and grain boundary effects.
\newblock {\em International Journal of solids and structures},
  41(18):5209--5230, 2004.

\bibitem{ozdemir2014modeling}
I.~\"Ozdemir and T.~Yal\c{c}inkaya.
\newblock Modeling of dislocation--grain boundary interactions in a strain
  gradient crystal plasticity framework.
\newblock {\em Computational Mechanics}, 54(2):255--268, 2014.

\bibitem{livingston1957multiple}
J.~Livingston and B.~Chalmers.
\newblock Multiple slip in bicrystal deformation.
\newblock {\em Acta Metallurgica}, 5(6):322--327, 1957.

\bibitem{luster1995compatibility}
J.~Luster and M.~Morris.
\newblock Compatibility of deformation in two-phase {Ti-Al} alloys:
  {D}ependence on microstructure and orientation relationships.
\newblock {\em Metallurgical and Materials Transactions A}, 26(7):1745--1756,
  1995.

\bibitem{bieler2014grain}
T.~Bieler, P.~Eisenlohr, C.~Zhang, H.~Phukan, and M.~Crimp.
\newblock Grain boundaries and interfaces in slip transfer.
\newblock {\em Current Opinion in Solid State and Materials Science},
  18(4):212--226, 2014.

\bibitem{Gottschalk2016443}
D.~Gottschalk, A.~McBride, B.~Reddy, A.~Javili, P.~Wriggers, and
  C.~Hirschberger.
\newblock Computational and theoretical aspects of a grain-boundary model that
  accounts for grain misorientation and grain-boundary orientation.
\newblock {\em Computational Materials Science}, 111:443 -- 459, 2016.

\bibitem{schmid1935kristallplastizitat}
E.~Schmid and W.~Boas.
\newblock {\em {K}ristallplastizit{\"a}t: {M}it besonderer
  {B}er{\"u}cksichtigung der {M}etalle}.
\newblock Springer Berlin Heidelberg, 1935.

\bibitem{eshelby1951xli}
J.~Eshelby, F.~Frank, and F.~Nabarro.
\newblock {XLI.} {T}he equilibrium of linear arrays of dislocations.
\newblock {\em The London, Edinburgh, and Dublin Philosophical Magazine and
  Journal of Science}, 42(327):351--364, 1951.

\bibitem{wo2004investigation}
P.~Wo and A.~Ngan.
\newblock Investigation of slip transmission behavior across grain boundaries
  in polycrystalline {Ni3Al} using nanoindentation.
\newblock {\em Journal of Materials Research}, 19(01):189--201, 2004.

\bibitem{lee1990tem}
T.~Lee, I.~Robertson, and H.~Birnbaum.
\newblock {TEM} in situ deformation study of the interaction of lattice
  dislocations with grain boundaries in metals.
\newblock {\em Philosophical Magazine A}, 62(1):131--153, 1990.

\bibitem{bamford1988thermodynamic}
T.~Bamford, W.~Clark, and R.~Wagoner.
\newblock A thermodynamic model of slip propagation.
\newblock {\em Scripta Metallurgica}, 22(12):1911--1916, 1988.

\bibitem{tiba2015incompatibility}
I.~Tiba, T.~Richeton, C.~Motz, H.~Vehoff, and S.~Berbenni.
\newblock Incompatibility stresses at grain boundaries in {Ni} bicrystalline
  micropillars analyzed by an anisotropic model and slip activity.
\newblock {\em Acta Materialia}, 83:227--238, 2015.

\bibitem{patriarca2013slip}
L.~Patriarca, W.~Abuzaid, H.~Sehitoglu, and H.~J. Maier.
\newblock Slip transmission in bcc {FeCr} polycrystal.
\newblock {\em Materials Science and Engineering: A}, 588:308--317, 2013.

\bibitem{werner1990slip}
E.~Werner and W.~Prantl.
\newblock Slip transfer across grain and phase boundaries.
\newblock {\em Acta Metallurgica et Materialia}, 38(3):533--537, 1990.

\bibitem{kumar2010influence}
B.~R. Kumar.
\newblock Influence of crystallographic textures on tensile properties of
  {316L} austenitic stainless steel.
\newblock {\em Journal of Materials Science}, 45(10):2598--2605, 2010.

\bibitem{demkowicz2011structure}
M.~Demkowicz and L.~Thilly.
\newblock Structure, shear resistance and interaction with point defects of
  interfaces in {Cu--Nb} nanocomposites synthesized by severe plastic
  deformation.
\newblock {\em Acta Materialia}, 59(20):7744--7756, 2011.

\bibitem{wang2011influence}
J.~Wang, R.~Hoagland, X.~Liu, and A.~Misra.
\newblock The influence of interface shear strength on the glide
  dislocation--interface interactions.
\newblock {\em Acta Materialia}, 59(8):3164--3173, 2011.

\bibitem{wang2012structure}
J.~Wang, K.~Kang, R.~Zhang, S.~Zheng, I.~Beyerlein, and N.~Mara.
\newblock Structure and property of interfaces in {ARB Cu/Nb} laminated
  composites.
\newblock {\em JOM}, 64(10):1208--1217, 2012.

\bibitem{spearot2014insights}
D.~E. Spearot and M.~D. Sangid.
\newblock Insights on slip transmission at grain boundaries from atomistic
  simulations.
\newblock {\em Current Opinion in Solid State and Materials Science},
  18(4):188--195, 2014.

\bibitem{bieler2009role}
T.~Bieler, P.~Eisenlohr, F.~Roters, D.~Kumar, D.~Mason, M.~Crimp, and D.~Raabe.
\newblock The role of heterogeneous deformation on damage nucleation at grain
  boundaries in single phase metals.
\newblock {\em International Journal of Plasticity}, 25(9):1655--1683, 2009.

\bibitem{sangid2012energetics}
M.~D. Sangid, T.~Ezaz, and H.~Sehitoglu.
\newblock Energetics of residual dislocations associated with slip--twin and
  slip--{GB}s interactions.
\newblock {\em Materials Science and Engineering: A}, 542:21--30, 2012.

\bibitem{dewald2011multiscale}
M.~Dewald and W.~Curtin.
\newblock Multiscale modeling of dislocation/grain-boundary interactions:
  {III}. 60$^\circ$ dislocations impinging on {$\Sigma$}3, {$\Sigma$}9 and
  {$\Sigma$}11 tilt boundaries in {Al}.
\newblock {\em Modelling and Simulation in Materials Science and Engineering},
  19(5):055002, 2011.

\bibitem{ezaz2011energy}
T.~Ezaz, M.~D. Sangid, and H.~Sehitoglu.
\newblock Energy barriers associated with slip--twin interactions.
\newblock {\em Philosophical Magazine}, 91(10):1464--1488, 2011.

\bibitem{zikry1996inelastic}
M.~Zikry and M.~Kao.
\newblock Inelastic microstructural failure mechanisms in crystalline materials
  with high angle grain boundaries.
\newblock {\em Journal of the Mechanics and Physics of Solids},
  44(11):1765--1798, 1996.

\bibitem{ekh2011influence}
M.~Ekh, S.~Bargmann, and M.~Grymer.
\newblock Influence of grain boundary conditions on modeling of size-dependence
  in polycrystals.
\newblock {\em Acta Mechanica}, 218(1-2):103--113, 2011.

\bibitem{shi2009grain}
J.~Shi and M.~Zikry.
\newblock Grain--boundary interactions and orientation effects on crack
  behavior in polycrystalline aggregates.
\newblock {\em International Journal of Solids and Structures},
  46(21):3914--3925, 2009.

\bibitem{shi2011modeling}
J.~Shi and M.~A. Zikry.
\newblock Modeling of grain boundary transmission, emission, absorption and
  overall crystalline behavior in {$\Sigma$}1, {$\Sigma$}3, and {$\Sigma$}17b
  bicrystals.
\newblock {\em Journal of Materials Research}, 26(14):1676--1687, 2011.

\bibitem{ma2006consideration}
A.~Ma, F.~Roters, and D.~Raabe.
\newblock On the consideration of interactions between dislocations and grain
  boundaries in crystal plasticity finite element modeling--theory,
  experiments, and simulations.
\newblock {\em Acta Materialia}, 54(8):2181--2194, 2006.

\bibitem{ashmawi2002prediction}
W.~Ashmawi and M.~Zikry.
\newblock Prediction of grain-boundary interfacial mechanisms in
  polycrystalline materials.
\newblock {\em Journal of Engineering Materials and Technology}, 124(1):88--96,
  2002.

\bibitem{mesarovic2010plasticity}
S.~D. Mesarovic.
\newblock Plasticity of crystals and interfaces: {F}rom discrete dislocations
  to size-dependent continuum theory.
\newblock {\em Theoretical and Applied Mechanics}, 37(4):289--332, 2010.

\bibitem{gurtin2005boundary}
M.~E. Gurtin and A.~Needleman.
\newblock Boundary conditions in small-deformation, single-crystal plasticity
  that account for the {B}urgers vector.
\newblock {\em Journal of the Mechanics and Physics of Solids}, 53(1):1--31,
  2005.

\bibitem{jin2008interactions}
Z.-H. Jin, P.~Gumbsch, K.~Albe, E.~Ma, K.~Lu, H.~Gleiter, and H.~Hahn.
\newblock Interactions between non-screw lattice dislocations and coherent twin
  boundaries in face-centered cubic metals.
\newblock {\em Acta Materialia}, 56(5):1126--1135, 2008.

\bibitem{cheng2008atomistic}
Y.~Cheng, M.~Mrovec, and P.~Gumbsch.
\newblock Atomistic simulations of interactions between the 1/2\textless
  111\textgreater edge dislocation and symmetric tilt grain boundaries in
  tungsten.
\newblock {\em Philosophical Magazine}, 88(4):547--560, 2008.

\bibitem{gibson2002slip}
M.~Gibson and C.~Forwood.
\newblock Slip transfer of deformation twins in duplex $\gamma$-based {Ti-Al}
  alloys: {Part III}. {T}ransfer across general large-angle $\gamma$-$\gamma$
  grain boundaries.
\newblock {\em Philosophical Magazine A}, 82(7):1381--1404, 2002.

\bibitem{soer2005incipient}
W.~Soer, K.~Aifantis, and J.~T.~M. De~Hosson.
\newblock Incipient plasticity during nanoindentation at grain boundaries in
  body-centered cubic metals.
\newblock {\em Acta Materialia}, 53(17):4665--4676, 2005.

\bibitem{liu1995dislocation}
F.~Liu, I.~Baker, and M.~Dudley.
\newblock Dislocation-grain boundary interactions in ice crystals.
\newblock {\em Philosophical Magazine A}, 71(1):15--42, 1995.

\end{thebibliography}

\end{document}